\documentclass[fleqn,usenatbib]{mnras}

\usepackage{newtxtext,newtxmath}

\usepackage[T1]{fontenc}
\usepackage{ae,aecompl}

\usepackage{graphicx}   
\usepackage{amsmath}    
\usepackage{amssymb}    

\usepackage{breakurl}

\newcommand{\sunlum}{~L$_{\odot}$~}
\newcommand{\kms}{~km s$^{-1}$~}
\newcommand{\kmsE}{~km s$^{-1}$}
\newcommand{\WR}{WR~147~}
\newcommand{\WRE}{WR~147}
\newcommand{\GTC}{{\sc GTC~}}
\newcommand{\GTCE}{{\sc GTC}}
\newcommand{\HST}{{\sc HST~}}
\newcommand{\HSTE}{{\sc HST}}
\newcommand{\dotMyr}{~M$_{\odot}$~yr$^{-1}$~}

\newcommand{\Chandra}{{\it Chandra~}}
\newcommand{\ChandraE}{{\it Chandra}}

\newcommand{\Gaia}{{\it Gaia~}}

\newcommand{\cmfgen}{{\sc cmfgen~}}
\newcommand{\cmfgenE}{{\sc cmfgen}}

\newcommand{\iraf}{{\sc iraf~}}

\newcommand{\Tstar}{T$_*$~}
\newcommand{\TstarE}{T$_*$}
\newcommand{\logL}{$\log$~L~}
\newcommand{\logLE}{$\log$~L}
\newcommand{\dotM}{$\log~\dot{M}$~}
\newcommand{\dotME}{$\log~\dot{M}$}

\newcommand{\rr}{\tilde{r}}
\newcommand{\vv}{\tilde{\varv}}

\title[Modelling of \WRE]
{A global view on the colliding-wind binary \WR\thanks{Based on 
observations with Gran Telescopio CANARIAS (programme ID
GTC93-17B; PI: P.Pessev).}
}

\author[S.A.Zhekov et al.]{Svetozar A. Zhekov$^1$\thanks{
E-mail: szhekov@astro.bas.bg}, Blagovest V. Petrov$^1$,
Toma V. Tomov$^2$\thanks{deceased, 2019 August 9} and Peter Pessev$^3$\\
$^1$Institute of Astronomy and National Astronomical Observatory
(Bulgarian Academy of Sciences),\\
72 Tsarigradsko Chaussee Blvd., Sofia 1784, Bulgaria\\
$^2$Centre for Astronomy, Faculty of Physics, Astronomy and
Informatics, Nicolaus Copernicus University, \\
Grudziadzka 5, 87-100 Torun, Poland\\
$^3$Instituto de Astrofisica de Canarias, 38200, La Laguna, Tenerife, Spain;\\ 
Departamento de Astrofisica, Universidad de La Laguna, 38206, La Laguna, 
Tenerife, Spain
}

\date{}

\pubyear{2019}

\begin{document}
\label{firstpage}
\pagerange{\pageref{firstpage}--\pageref{lastpage}}
\maketitle

\begin{abstract}
We present results from  a global view on the colliding-wind binary
\WRE. We analysed new optical spectra of \WR obtained with Gran 
Telescopio CANARIAS and archive spectra from the Hubble Space 
Telescope by making use of modern atmosphere models 
accounting for optically thin clumping. 
We adopted a grid-modelling approach to derive some basic physical 
characteristics of both stellar components in \WRE.
For the currently accepted distance of 630 pc to \WRE, the values of 
mass-loss rate derived from modelling its optical spectra are in 
acceptable correspondence with that from modelling its X-ray emission.
However, they give a lower radio flux than observed.  A plausible 
solution for this problem could be if the volume filling factor at 
large distances from the star (radio-formation region) is smaller
than close to the star (optical-formation region). Adopting this, the
model can match well {\it both} optical and thermal radio 
emission from \WRE.
The global view on the colliding-wind
binary \WR thus shows that its 
observational properties in different spectral domains 
can be explained in a self-consistent physical picture.

\end{abstract}

\begin{keywords}
shock waves --- stars: individual: \WR --- stars: Wolf-Rayet ---
stars: winds, outflows.
\end{keywords}



\section{Introduction}
Massive stars of early type, Wolf-Rayet (WR) and OB, possess massive
and fast supersonic stellar winds: typically $\dot{M} \sim 10^{-5}$ 
(WR) and $\sim 10^{-7} - 10^{-6}$ (OB)\dotMyr and 
$V_{wind} = 1000 - 5000$\kmsE.
If both components of a binary system are massive stars, their
powerful winds will interact resulting in enhanced X-ray emission,
produced in the colliding-stellar-wind (CSW) shocks, as first proposed
by \citet{pri_us_76} and \citet{cherep_76}.
However, the CSW role is not limited to the X-ray emission of massive 
binaries. For example, wide WR+OB binaries are non-thermal radio 
sources \citep{do_wi_00} and this emission is believed to originate
from their CSW region. Also, CSWs are invoked to explain the recurrent
infrared bursts observed in the so called episodic dust makers, which
are wide WR+O binaries with high orbital eccentricity
(\citealt{williams_95}, \citealt{williams_08} and references therein).

We note that the basic parameters that determine the physical
characteristics of the shocked CSW plasma are the mass-loss rate and
velocity of the stellar winds and the binary separation
(\citealt{leb_myas_90}; \citealt{luo_90}; \citealt{stevens_92};
\citealt{mzh_93}).
Since the bulk emission of the CSW plasma is in X-rays, comparison of
observed X-ray spectra of wide CSW binaries with theoretical
predictions could provide additional constraints on the stellar wind
properties of the binary components. 

Interestingly, the {\it direct} modelling of the X-ray emission from 
the wide WR+O binaries WR~137, WR~146 and WR~147 in the framework of 
the CSW picture required mass-loss rate lower than the currently
accepted values for these objects: of about one order of magnitude 
for WR~137 and WR~146, and of about a factor of four for WR~147
(\citealt{zh_15}; \citealt{zh_17}; \citealt{zhp_10b}). Similarly,
reduced mass-loss rates of the stellar components by a factor of 
$\sim 7.5-7.7$ was deduced from a direct modelling of the X-ray 
emission of the O$+$O binary Cyg OB2 9 (Schulte 9;
\citealt{parkin_14}).

In order to find a solution to this problem, a global view on the
spectral modelling was proposed (for details see section 5 in 
\citealt{zh_17}). Namely, we need to carry out spectral modelling of
the observational properties of massive stars in different spectral
domains (e.g., radio, optical/UV, X-ray) to see whether we could 
reconcile the results on mass-loss rates from different analyses.

In this study, we chose to adopt a global view on the WR+OB binary
\WRE: it is a classical CSW binary; it has been studied in considerable 
detail almost along the entire spectrum; it has the smallest mass-loss 
mismatch amongst the studied CSW binaries so far (see above).
It is worth noting that \WR has been spatially resolved in each
spectral domain: radio \citep{moran_89}, infrared \citep{williams_97},
optical \citep{niemela_98}, X-rays \citep{zhp_10a}. This could 
potentially allow for obtaining observational properties of both 
stellar components in this CSW binary: a WN8 Wolf-Rayet star and an
OB star.

Our paper is organized as follows.
In Section~\ref{sec:data}, we describe the new and archival optical
observations of \WRE.
In Section~\ref{sec:modelling}, we give details about the 
modelling of the optical spectra of both binary components and the corresponding results.
In Section~\ref{sec:discussion}, we discuss our results, and we 
present our conclusions in Section~\ref{sec:conclusions}.

\begin{figure*}
\begin{center}
\includegraphics[width=2.3in, height=1.64in]{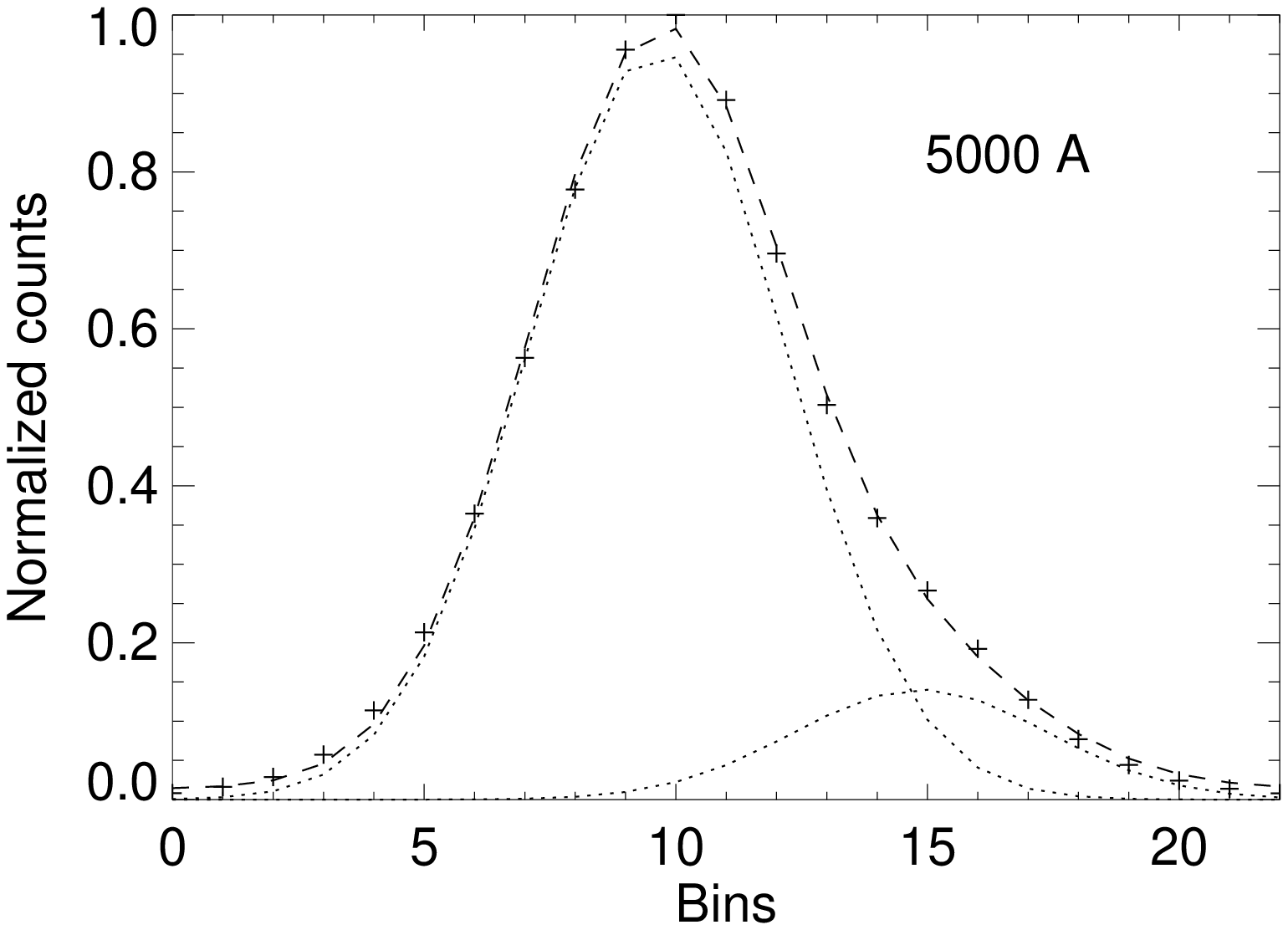}
\includegraphics[width=2.3in, height=1.64in]{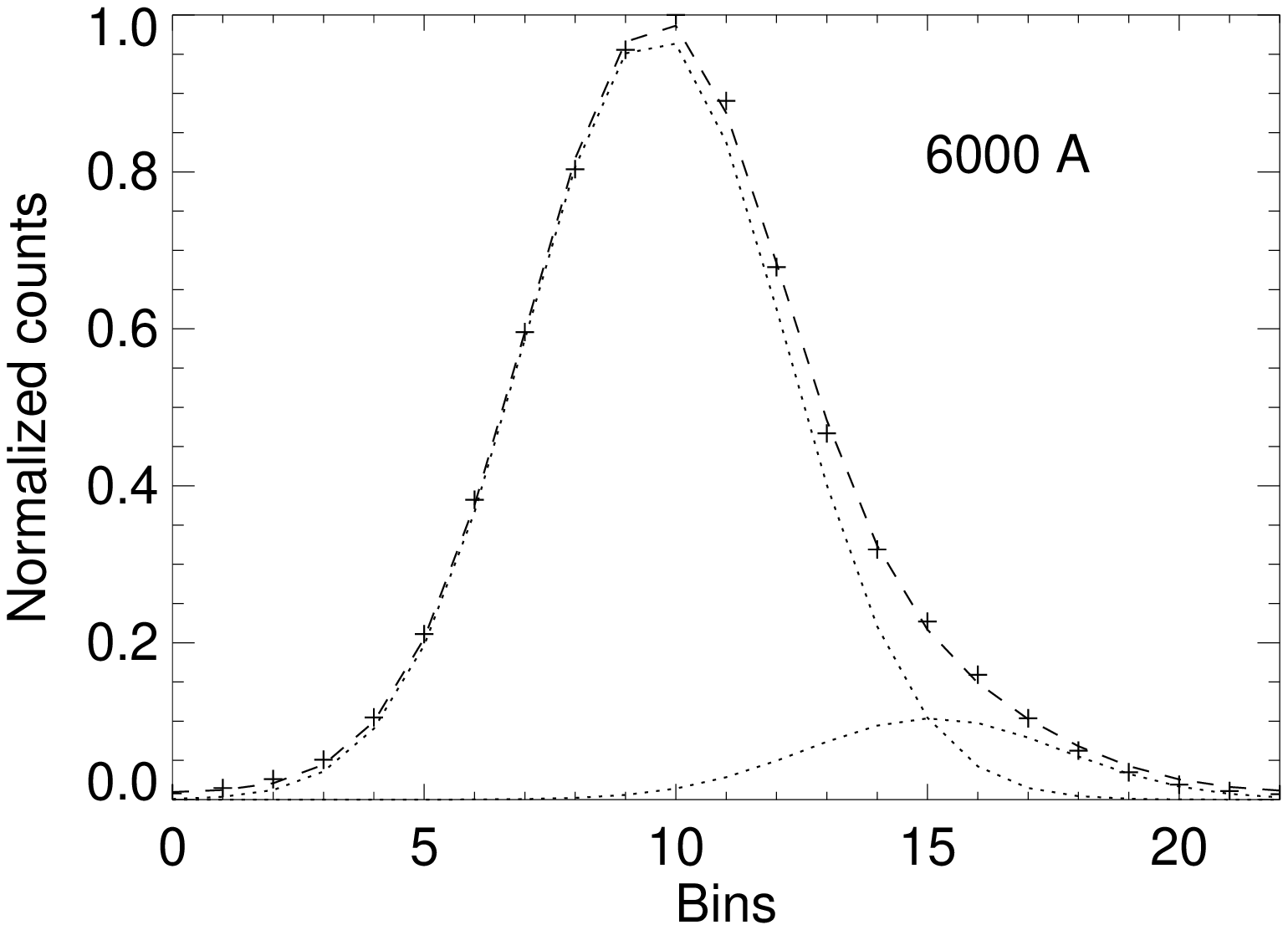}
\includegraphics[width=2.3in, height=1.64in]{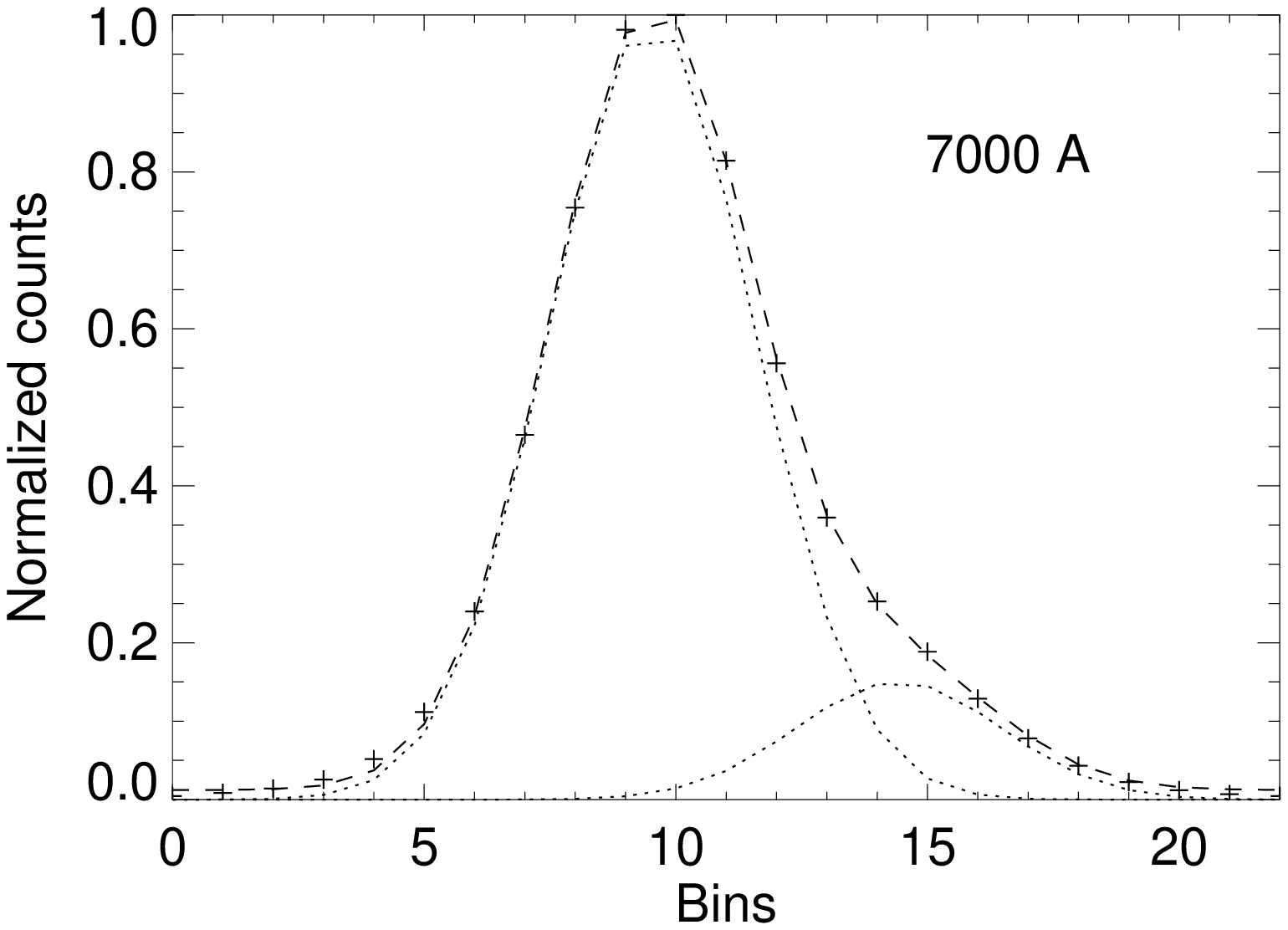}
\includegraphics[width=2.3in, height=1.64in]{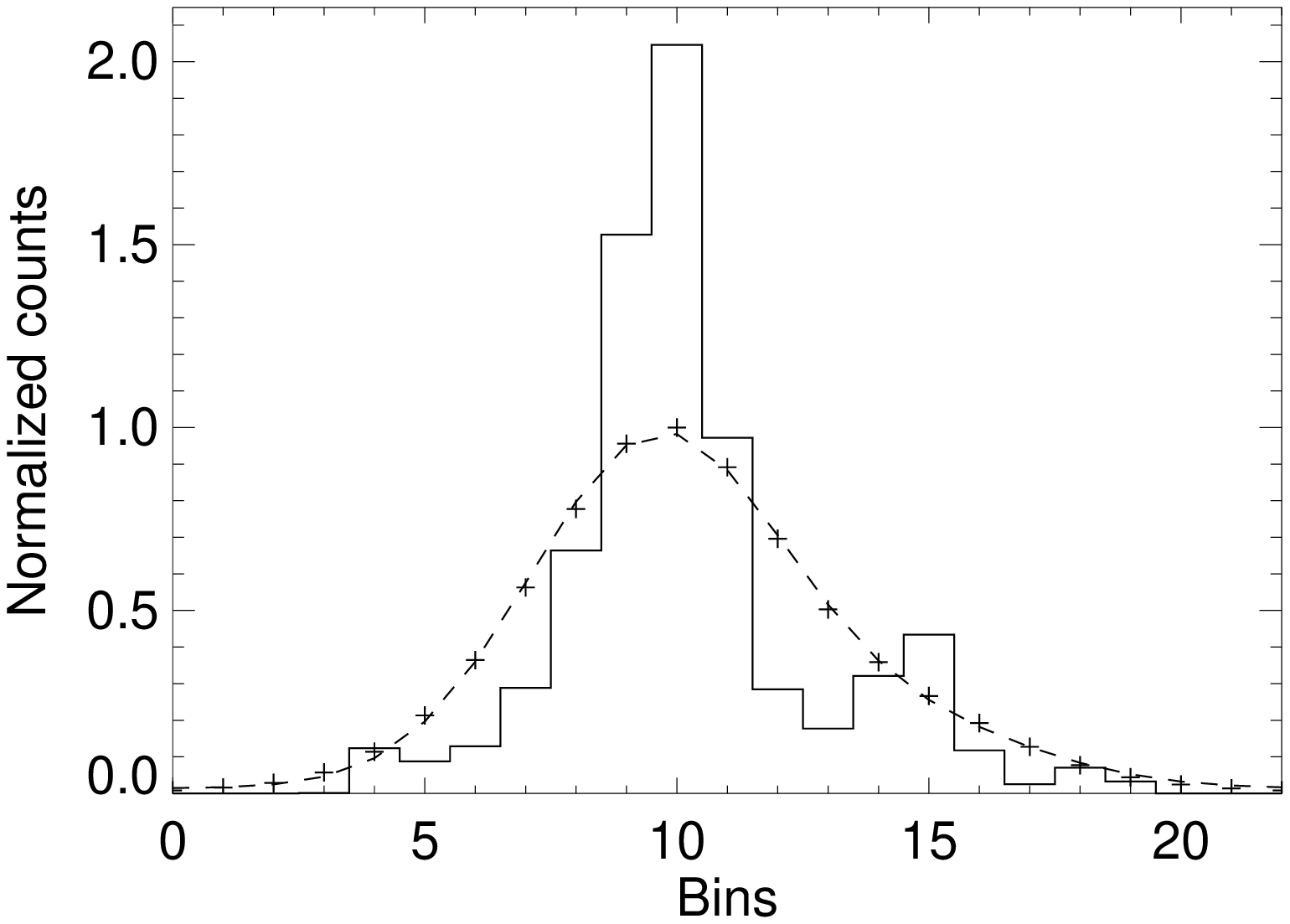}
\includegraphics[width=2.3in, height=1.64in]{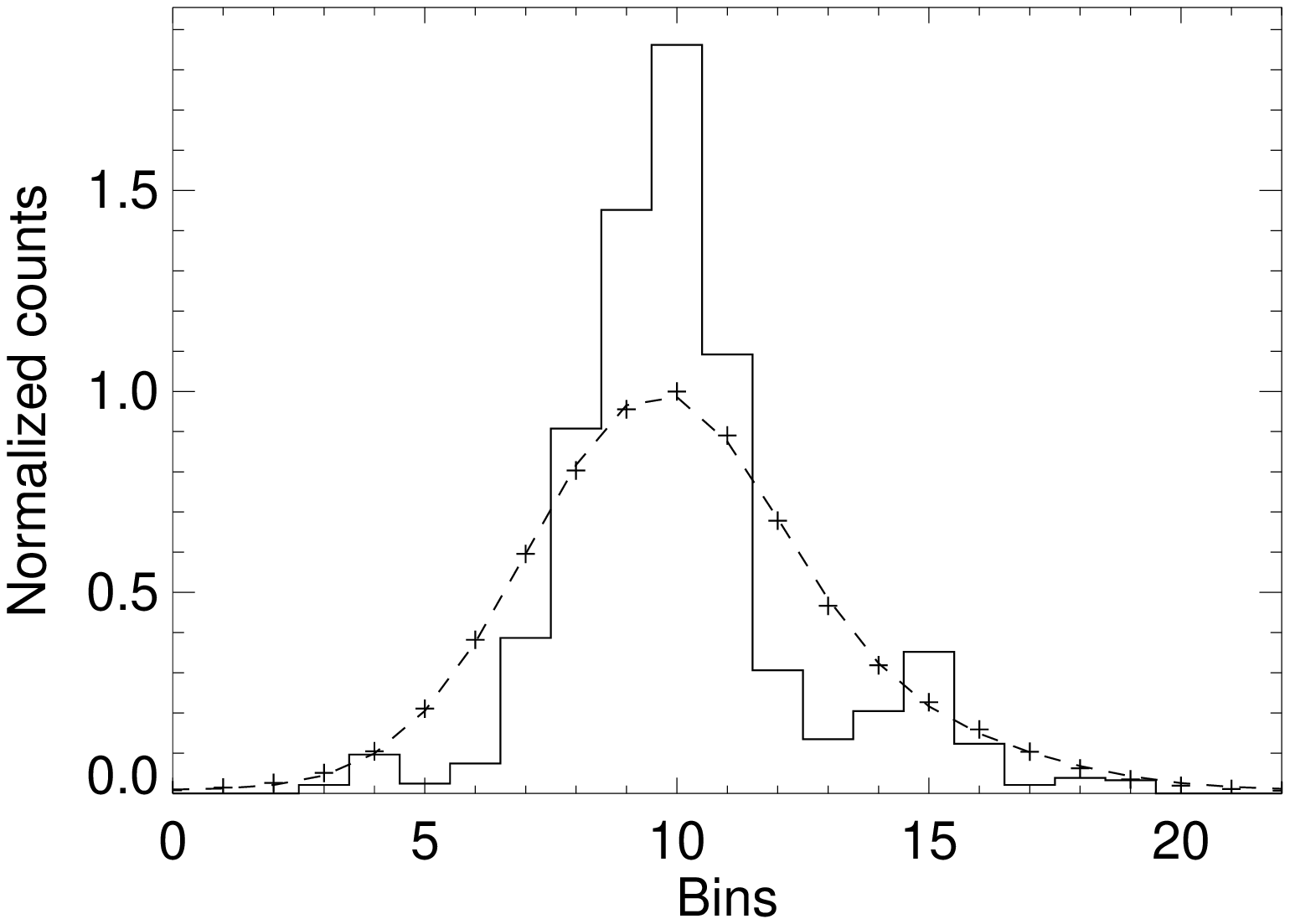}
\includegraphics[width=2.3in, height=1.64in]{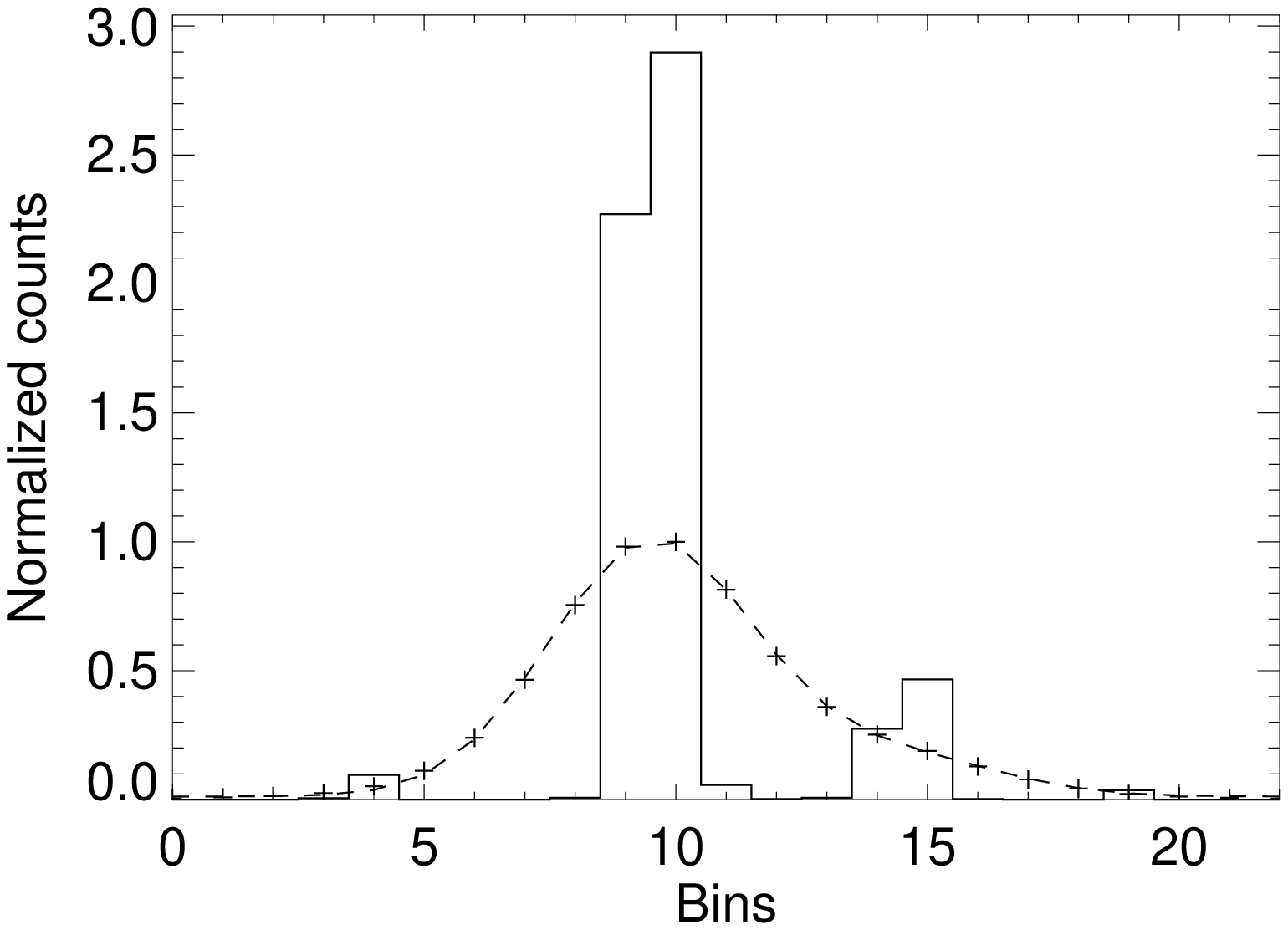}
\end{center}
\caption{Examples of the spectral `deconvolution'.
The two-Gaussian fit results are shown in the first row.
The data are depicted with `+' sign and dashed line; the fit
components are shown with dotted lines.
Results from the Richardson-Lucy deconvolution (solid line) are shown
in the second row .
The three columns (from {\it left} to {\it right} show the
cross-dispersion cuts at $\lambda = 5000$~\AA, $\lambda = 6000$~\AA
\, and $\lambda = 7000$~\AA, respectively.
}
\label{fig:deconv}
\end{figure*}

\begin{figure}
\begin{center}
\includegraphics[width=\columnwidth]{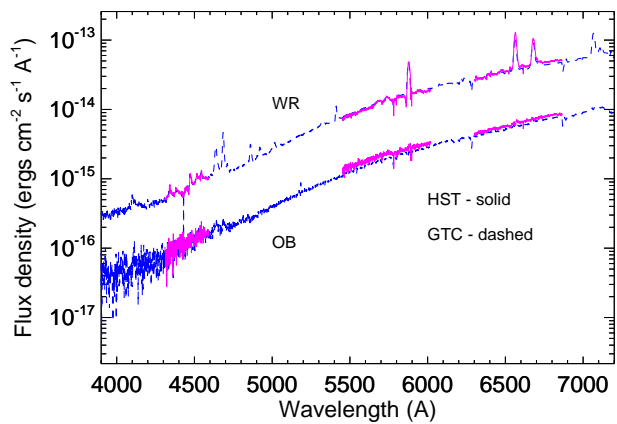}
\end{center}
\caption{The HST vs. GTC comparison of the mean spectra of the WR
and OB components in \WRE. The HST and GTC spectra are in magenta and 
blue color, respectively.
}
\label{fig:hst_vs_gtc}
\end{figure}

\section{Observations and data reduction}
\label{sec:data}

\subsection{Observations with Gran Telescopio CANARIAS}
\label{sec:data_gtc}

\WR was observed with Gran Telescopio CANARIAS 
on 2017 September 18 under excellent 
photometric conditions (clear skies and seeing of $\approx 0\farcs6$;
see also Section~\ref{sec:data_gtc_dcnv}).  The long-slit mode of the 
OSIRIS spectrograph\footnote{For details see \url{
http://www.gtc.iac.es/instruments/osiris/\#Longslit_Spectroscopy}}
 was used in combination
with the R1000B grism that provides spectra in the 3630 - 7500 \AA~
range with a resolution of $\lambda / \Delta \lambda \approx 1018$. 
The slit width was $0\farcs6$ and the slit length was aligned with the 
position angle of the \WR binary (PA $= -10\degr$; see table 2 in 
 \citealt{niemela_98}).
Twelve scientific exposures of \WR were carried out with this OSIRIS
setup. However, two of them (exposure time of 180 and 360 s) resulted
in saturated spectra at wavelengths $>\,\sim 6000$~\AA. So, only ten 
spectra (each with a 90-s exposure time) were used in the current 
analysis.

Following the standard calibration procedure for OSIRIS, the
calibration frames were acquired at the end of the observing night,
utilizing the Naysmith B focal station Instrument Calibration Module
(ICM) and the same instrument configuration. This calibration approach
is feasible due to the proven stability of OSIRIS and has been
validated all through the instrument operation.

For the flux calibration, we used spectra of the spectrophotometric
standard star G191-B2B, obtained in the same night with the same
instrument configuration and under similar observing conditions
through a $2\farcs5$ wide long-slit. Thus assuring that all the
flux from the standard star has passed through the observing system 
and reached the instrument detector.

For the flat field corrections, wavelength
calibration and correcting the distortion,
we used the standard \iraf\footnote{\iraf is          
distributed by the National Optical Astronomy Observatories, which are
operated by the Association of Universities for Research in Astronomy,
Inc., under cooperative agreement with the National Science
Foundation.} tasks in the {\sc twospec} package.
We note that individual spectra of the WR and OB components in \WR 
were extracted first (see Section~\ref{sec:data_gtc_dcnv}) and then
flux-calibrated.

\subsubsection{Spectral `deconvolution'}
\label{sec:data_gtc_dcnv}

Given the brightness difference between the WR and OB components in 
\WR of about 2 mag in the V-filter and the measured angular separation 
of $0\farcs64$ \citep{niemela_98}, we opted out for the $1\times1$ 
binning detector readout mode, instead of the nominal $2\times2$ 
binning of OSIRIS. Nevertheless, the resulting pixel size is still 
$0\farcs125$ and we had to adopt some `deconvolution' technique aimed 
at extracting a separate spectrum for the WR and OB component, 
respectively.

As a first step, we performed a 'spatial' fit (i.e., in the
cross-dispersion direction) at each `spectral' bin (wavelength) of 
the GTC-OSIRIS data of \WRE. The model function consisted of two 
Gaussian components and a continuum. The fit parameters were total 
flux and position of each component, the Gaussian $\sigma$ (being the 
same for both components) and the continuum level.

To check our `deconvolution' approach, the Richardson-Lucy algorithm
was used as well. It revealed that two components {\it do } exist in 
the OSIRIS data in the cross-dispersion direction. 

Interestingly, the results from our two-component fits and the
Richardson-Lucy deconvolution (\citealt{richardson_72};
\citealt{lucy_74}) showed that we do find two spectral
components that are separated at $\sim$ 0\farcs6 - 0\farcs7 in
`spatial' (cross-dispersion) direction and the `southern' component
(WR star) is brighter. Just to recall that the separation of the two
stellar components in \WR is 0\farcs64$\pm$0\farcs16
\citep{niemela_98}. Figure~\ref{fig:deconv} presents some results from 
our deconvolution experiment.

All this gave us confidence to proceed with the two-Gaussian spectral
`deconvolution' of the GTC-OSIRIS data of \WRE. A fixed separation
between the two components of 0\farcs64 was adopted in these fits.
Thus, the GTC-OSIRIS spectra  were subject to our 
`deconvolution' procedure and the WR and OB spectra were extracted for 
each data set. To increase the signal-to-noise in the spectra, all 
the 90-sec-exposure spectra were combined which resulted in a mean 
spectrum for each stellar component in \WRE.

Finally, the mean GTC spectrum of each stellar component was corrected
for the loss of flux that fell outside the spectrograph slit. This was
done by adopting a 2-D Gaussian PSF (point-spread function) and 
estimating the fraction of the
seeing that fell in the slit. An average seeing of 
0\farcs70$\pm$0\farcs03 (FWHM) was derived for the 90-sec data, which
is the mean value of the spatial component shape as derived from our
`deconvolution' procedure for each data set 
(for comparison, the Differential Image Motion Monitor at the Las
Moradas site located 300 meters East from the GTC, also known as
IAC-DIMM, gave an average seeing of 0\farcs74$\pm$0\farcs04  for the 
same night).
This estimate 
showed that a fraction of 0.844 of the total flux from \WR fell into 
the spectrograph slit, therefore, a correction coefficient of 1.184 
was adopted for the GTC spectra of \WRE. 

Figure~\ref{fig:hst_vs_gtc} shows how the GTC spectra of \WR compare
with those from the Hubble Space Telescope (HST). 
We note the good correspondence between the two
data sets and we underline that the HST data allow for a standard
spectral extraction of the WR and OB components (see
Section~\ref{sec:data_hst}), that is, {\it no }
`deconvolution' technique has been adopted in that case.
Thus, we feel confident to proceed further with the
spectral analysis of the GTC data on \WRE.

\subsection{Archive data from the Hubble Space Telescope}
\label{sec:data_hst}

For the purpose of this study, we made use of archival data from the
Hubble Space 
Telescope\footnote{See \url{https://mast.stsci.edu/portal/Mashup/Clients/Mast/Portal.html}
}
 that provide spatially resolved spectra
of both binary components in \WRE. The long-slit spectroscopy of \WR
was obtained with STIS (Space Telescope Imaging Spectrograph) in
combination with the G430M and G750M gratings (Obs IDs: o5d503010,
o5d503020 and o5d503040), which provide spectra in the
3020-5610~\AA~ ($\lambda / \Delta \lambda = 5390-10020$)
and 
5450-10140~\AA~ ($\lambda / \Delta \lambda = 4870-9050$)
range, respectively\footnote{
See
\url{http://www.stsci.edu/hst/instrumentation/stis/instrument-design/gratings--prism}
}.
These archival spectra are flux-calibrated and we note that they were
already discussed by \citet{lepine_01}. As noted by these authors,
the two stars in the \WR system are clearly resolved, which allows for
using standard \iraf aperture extraction to obtain their spectra. We 
thus adopted the same approach for the spectral extraction.

\section{Spectral modelling}
\label{sec:modelling}

\subsection{The stellar atmosphere model}
\label{sec:cmfgen}

For the purposes of our investigation, we employ the non-LTE radiative
transfer code \cmfgen (\citealt{hillier_99}; 
\citealt{hillier_01})\footnote{For more details and documentation on
\cmfgenE, see
\url{http://kookaburra.phyast.pitt.edu/hillier/web/CMFGEN.htm}},
which is a fully line-blanketed atmosphere code, designed to solve the
statistical equilibrium and radiative transfer equations in spherical
geometry.

The basic input parameters are stellar mass, luminosity, effective 
temperature, mass-loss rate, wind velocity and chemical composition 
(H, He, C, N, O, Ne, Mg, Al, Si, P, S, Ar, Ca, and Fe abundances; for
model atoms see Table~\ref{tab:atom} in Appendix~\ref{app_atom}). 
Stellar effective temperature in \cmfgen is defined using the 
Stefan-Boltzmann law with reference radius, $R_{2/3}$, specified 
for the Rosseland opacity  of 2/3: 
\Tstar $= (L / 4 \pi R^2_{2/3})^{1/4} $, 
where $L$ is the stellar luminosity.

Observations indicate that stellar winds of massive stars are not 
homogeneous, instead the stellar winds contain inhomogeneities or 
clumps. These clumps affect the stellar spectra and therefore 
modelling non-homogeneous stellar atmospheres is required.

\cmfgen can take optically thin clumping (microclumping) into account.
Microclumping approach is based on the hypothesis that the wind
consists of `clumps', which have enhanced density and dimension 
smaller than the photon mean free path. The density of the clumps is 
enhanced by a clumping factor $D$, compared to the mean density of 
the wind. This factor is described by the volume filling factor with 
relation $f=1/D$, assuming that the inter-clump medium is void. Our 
models are calculated with distant variable volume filing factor 
$f_{\infty} = 0.1$, prescribed by the following law (see eq. 5 in
\citealt{hillier_99}):
\begin{equation}
   f(r) = f_{\infty} + (1 - f_{\infty})\exp[-\varv(r)/\varv_{cl}]
   \label{eqn:clumping}
\end{equation}
where, $v_{cl}$ is the velocity at which clumping starts. We have 
chosen the clumping to start at $\varv_{cl} = 30$ \kmsE, which is near 
the sonic point.

It should be stated that \cmfgen does not currently solve the wind
momentum equation. Therefore, the wind velocity structure has to be
adopted.  In our models, the wind velocity structure is described by a
standard $\beta$-type velocity law with exponent $\beta = 1$.  The 
velocity law is joined to the hydrostatic part of the wind just below 
the sonic point, where the wind speed reaches local speed of sound.

The choice of having single values for the volume filling
factor and the velocity-law exponent is aimed at minimizing the number
of free parameters in the models (see
Section~\ref{sec:grid_modelling}). The specific choice of
$f_{\infty} = 0.1$ and $\beta = 1$ is based on a previous study
of the optical-IR spectra of \WR by \citet{morris_00}. 
These parameter values could be considered standard in spectral
 modelling of massive Wolf-Rayet stars (e.g., \citealt{hamann_06};
\citealt{sander_12}).

\subsection{Grid modelling}
\label{sec:grid_modelling}
We recall that the purpose of the spectral modelling by making use of
stellar atmosphere models (e.g., the \cmfgen code) is to derive some
basic physical characteristics of the studied massive stars: e.g., the
mass-loss rate ($\dot{M}$), luminosity (L), effective temperature
(\TstarE). 
It is well known, but it is nevertheless worth recalling that the 
spectrum of a massive star is determined by the ionization structure 
of its wind. In general, the ionization structure at a given point in 
the stellar wind depends on the ratio of the ionizing photons to the 
gas number density. The bulk of ionizing photons is determined by
both the stellar luminosity and the stellar temperature (controlling
the shape of the underlying continuum). On the other hand, the 
mass-loss rate of the massive star is the main factor that sets the
gas number density.

Therefore, we fit an observed spectrum {\it directly} 
(i.e., performing a full spectral comparison) in order to derive these
specific stellar parameters: $\dot{M}$, L, \TstarE.
Since the observed spectrum is 
flux-calibrated, we need to specify the distance to the studied object 
and to take into account the interstellar absorption. As a result, the 
fit provides the best value for the quantity E(B-V) adopting the 
standard ISM extinction curve ($R_V = 3.1$) of \citet{fitzpatrick_99}.
We next discuss a few important items for our fitting procedure.

(i) We note that all the spectral fits were done by making use of the
Levenberg-Marquardt method for non-linear $\chi^2$ fitting, which via
iterations finds the best fit (the minimum $\chi^2$ value) and 
requires an estimate of the theoretical function at each iteration 
(e.g., section~15.5 in \citealt{press_92}).
Since it is not feasible to directly calculate a theoretical 
(unabsorbed) spectrum for a given set of stellar parameters (too long
a computational time), the theoretical spectrum for each iteration is 
based on a preliminarily built grid of model spectra for a specific 
range of values of each of the stellar parameters of interest: i.e., 
\dotME, \logL and \TstarE. So, at each iteration we first find in 
which cell of the  3-D model grid the current parameter set (\dotME,
\logLE, \TstarE) is located. Then the resultant spectrum is estimated 
by interpolating (with specific weights = inverse distance-square) 
between all the model spectra which define that grid cell (eight 
spectra in total). This interpolation is similar to the Shepard's 
method \citep{shepard_68} but it is adopted {\it locally}:
$$
   Sp = \sum_{i = 1}^{8} w_i Sp_i, \hspace{1cm} 
   w_i = \frac{ \frac{1}{l_i^2}}{w_{tot}}, \hspace{1cm}
   w_{tot} = \sum_{i = 1}^{8} \frac{1}{l_i^2}
$$
where $l_i$ is the distance from the point (\dotME, \logLE, \TstarE) 
in the 3-D parameter space to each of the eight spectra ($Sp_i$) that
correspond to the vertexes of the current grid cell. We note that the
axes of the model grid are dimensionless and span the range  [0, 1],
that is each parameter is normalized to its range: $\bar{p} = (p -
p_{min}) / (p_{max} - p_{min})$, where $p$ is a given parameter and
$p_{min}, p_{max} $ are its minimum and maximum values, and $\bar{p}
\in [0, 1] $. Thus, all
the physical parameters of the model grid are `equal' in defining the 
distance in the 3-D parameter space.

(ii) We recall that the Levenberg-Marquardt method is a non-linear
$\chi^2$ fitting method and thus requires an initial guess of the
model parameters. Also,  due to the complex topology of the
$\chi^2$- surface of the non-linear methods very often a local minimum 
is found that may not be the true minimum we are looking for. To
circumvent this issue, we adopted the following approach in our grid 
modelling. 

We create a large number of initial parameter sets by randomly
choosing parameter values in the grid range for each parameter. For
each set of initial parameters, the Levenberg-Marquardt method finds
the corresponding `best fit', which is characterized by its $\chi^2$ 
value. We find the minimum value amongst all the $\chi^2$ values (a
great number of them, e.g., 10 000 or so), thus, starting from that
particular set of initial parameters, we find the parameters of the
spectral model that provide the best fit to the observed spectrum:
(\dotME, \logLE, \TstarE)$_{best}$.

(iii) As a final check, we calculate the theoretical model directly 
with \cmfgen by using exactly these `best-fit' parameters derived from 
the grid fitting and use it to fit the observed spectrum.

\begin{figure*}
\begin{center}
\includegraphics[width=2.8in, height=2.0in]{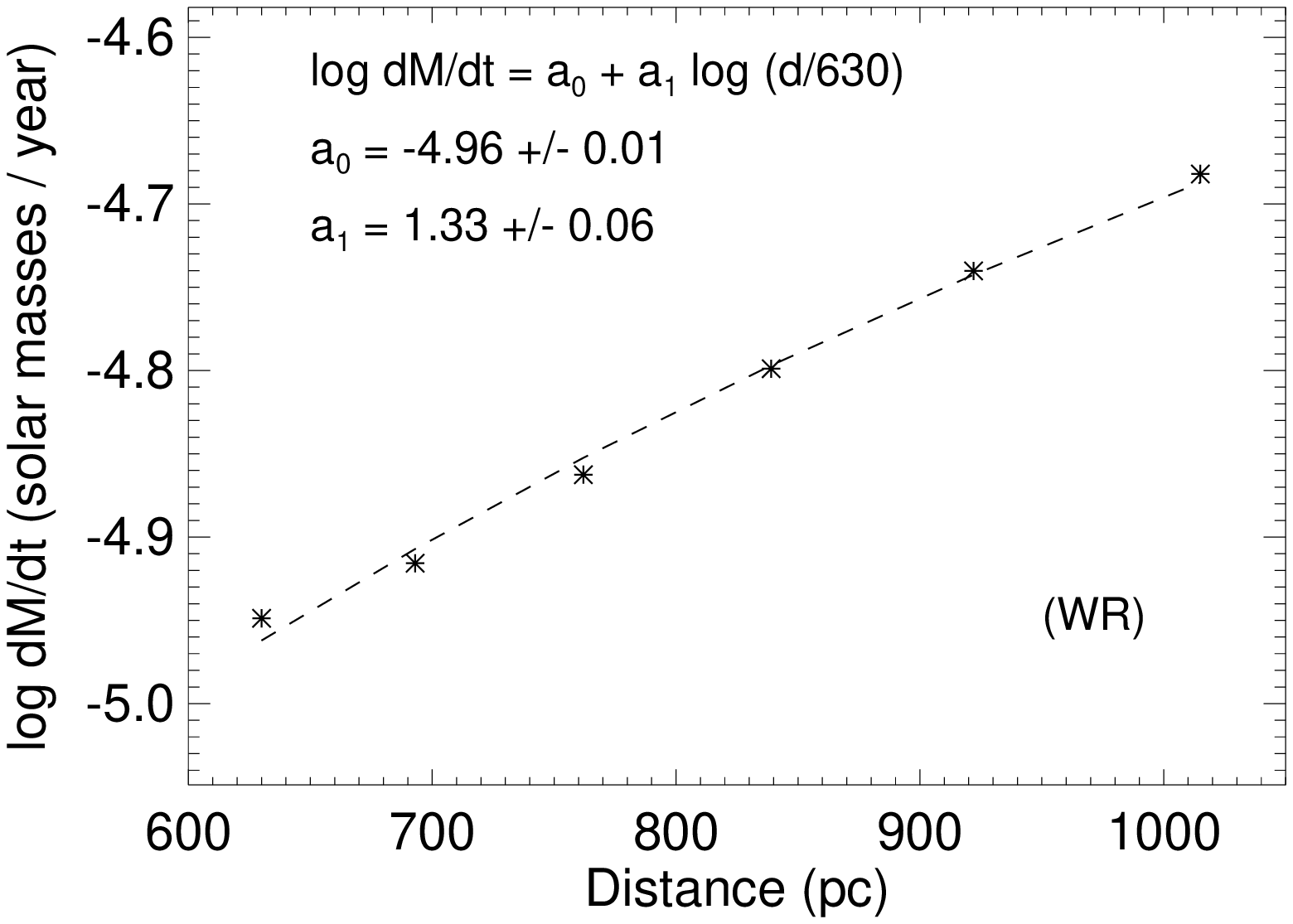}
\includegraphics[width=2.8in, height=2.0in]{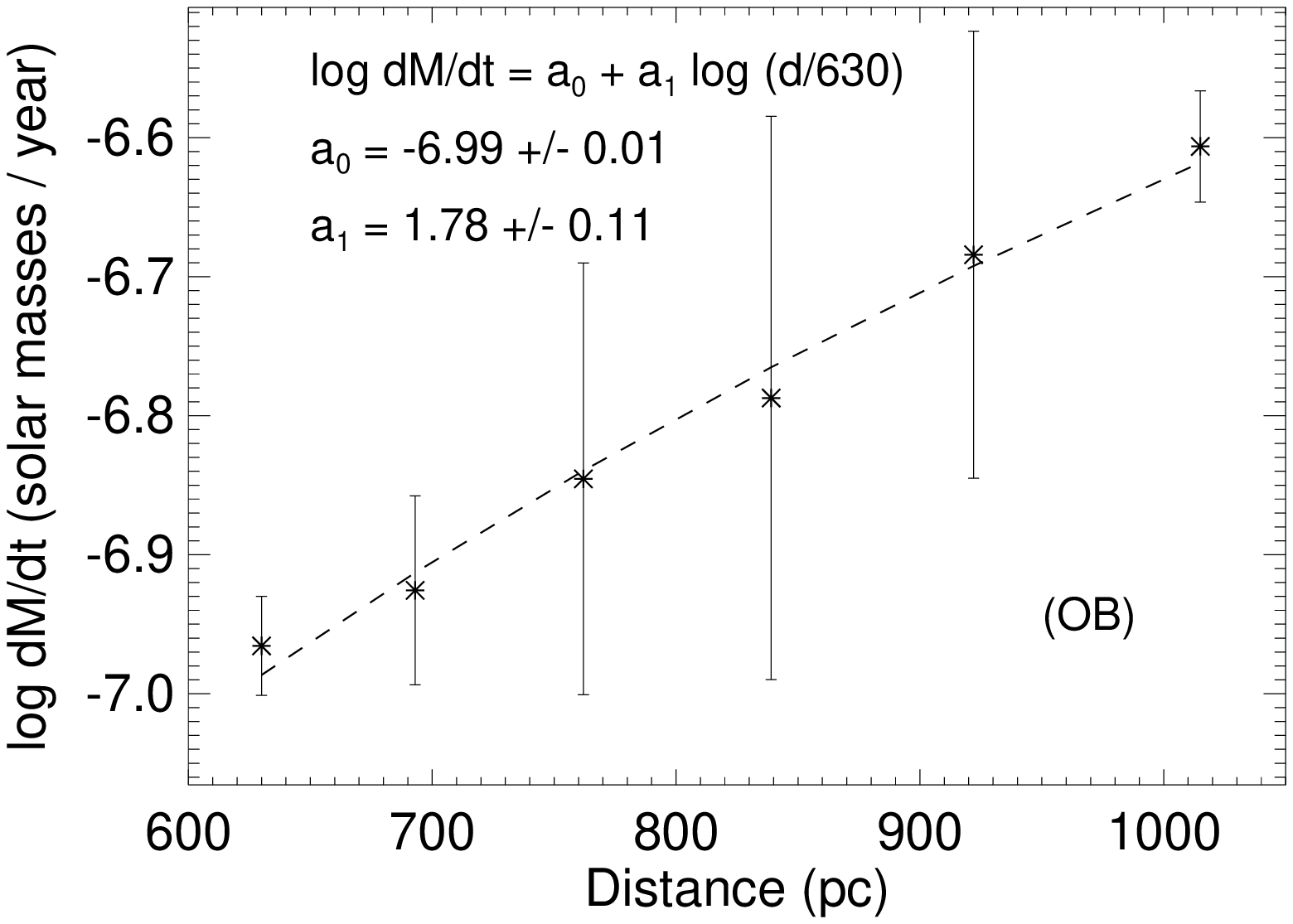}
\includegraphics[width=2.8in, height=2.0in]{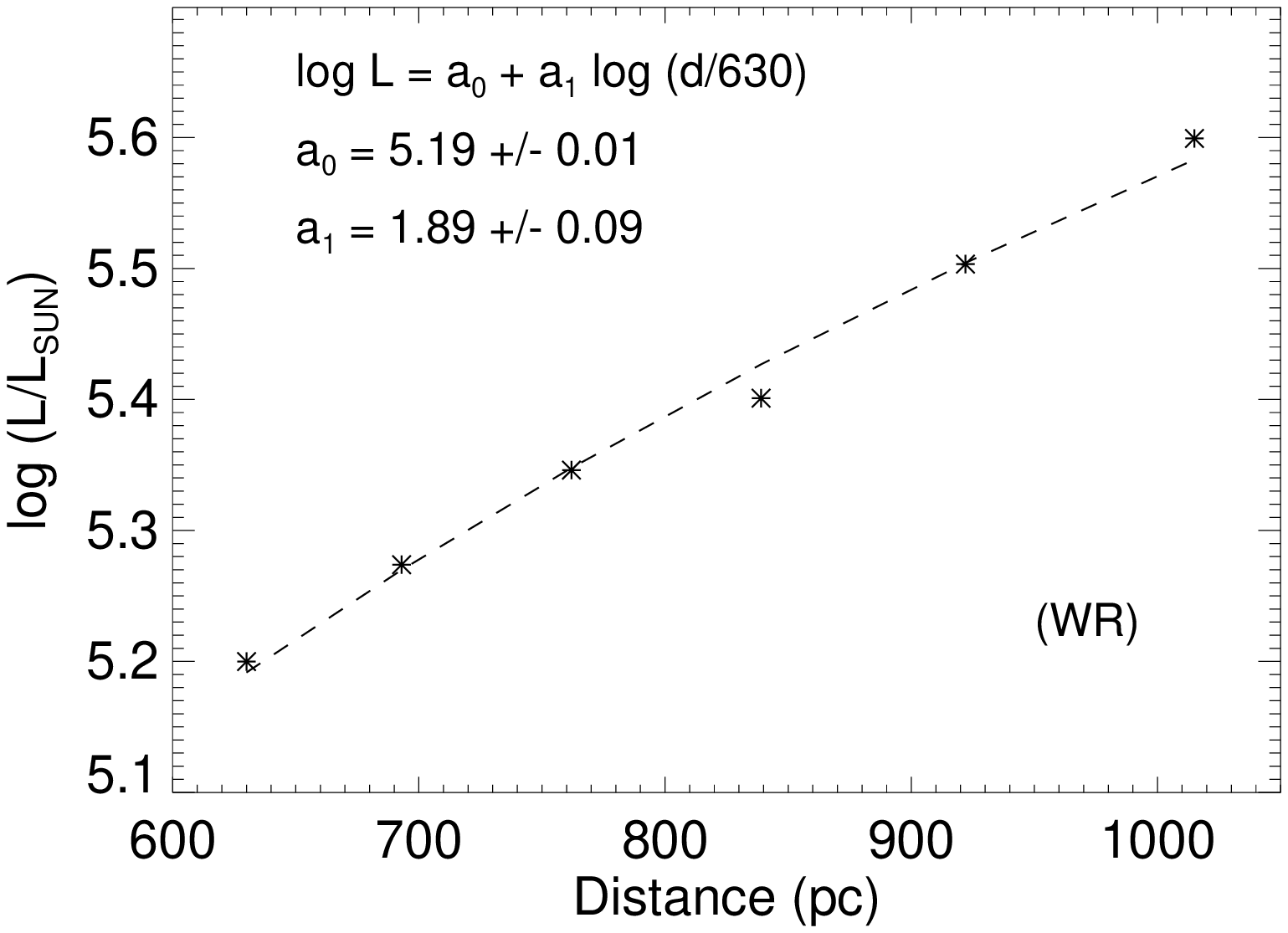}
\includegraphics[width=2.8in, height=2.0in]{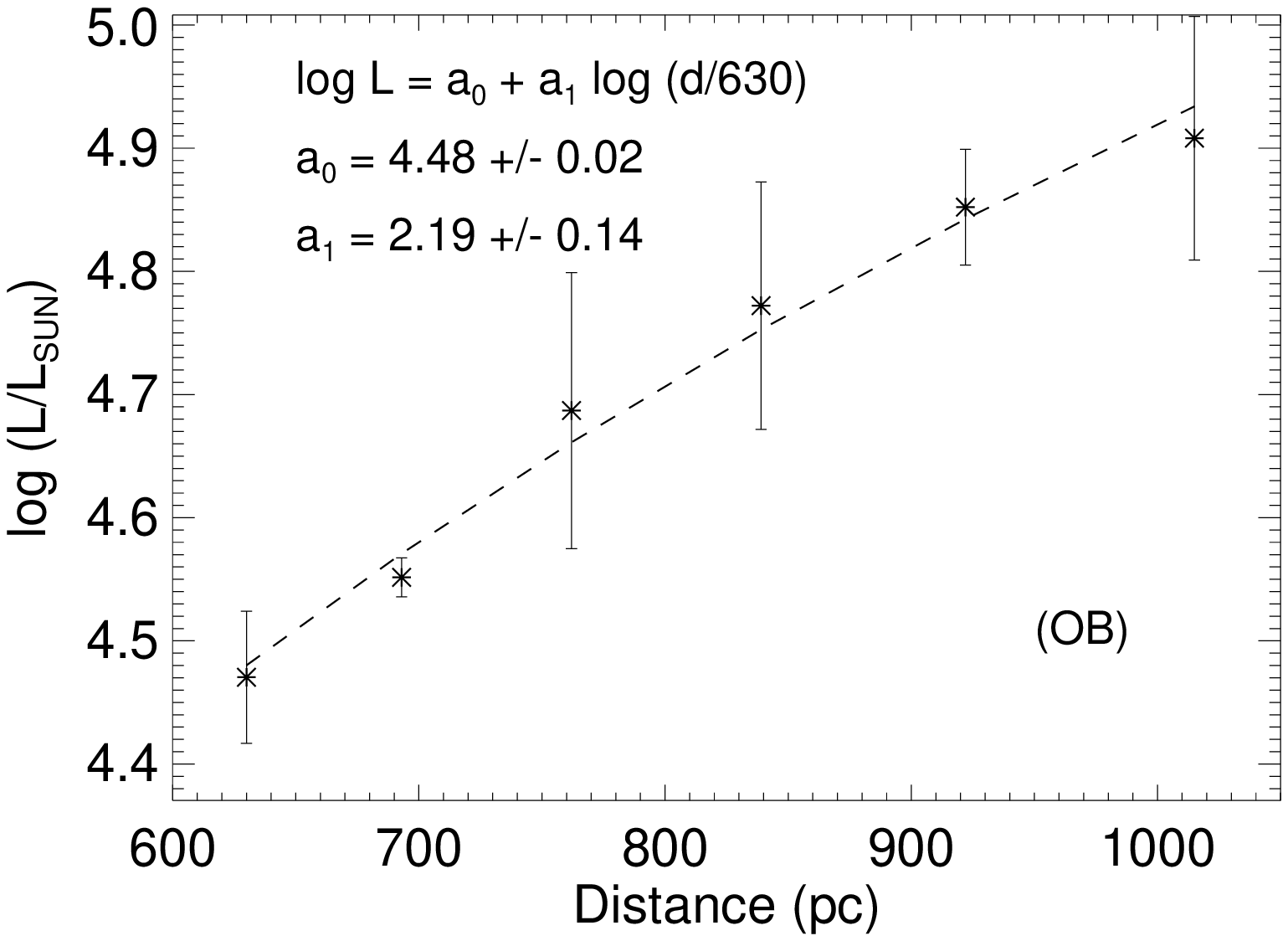}
\includegraphics[width=2.8in, height=2.0in]{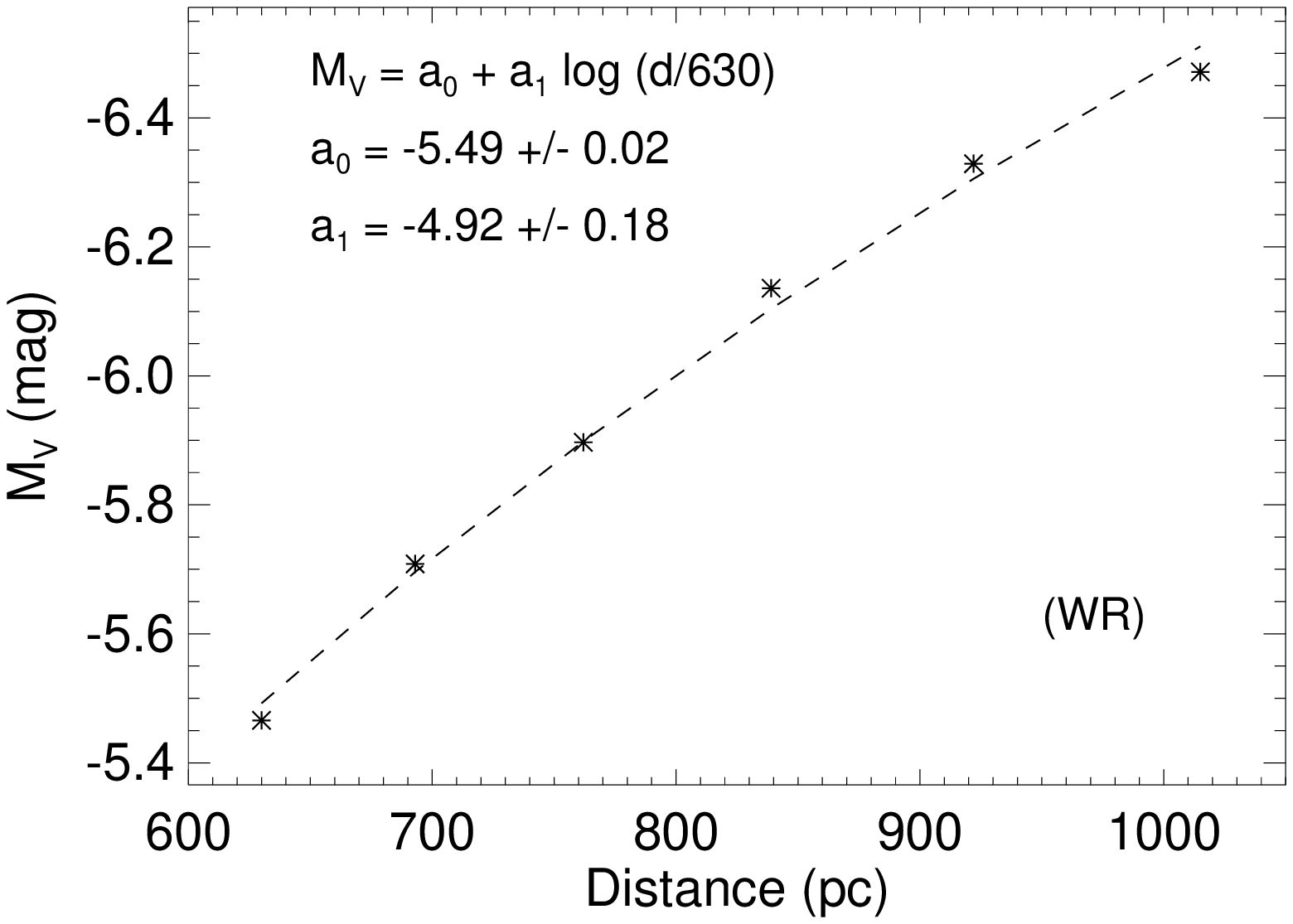}
\includegraphics[width=2.8in, height=2.0in]{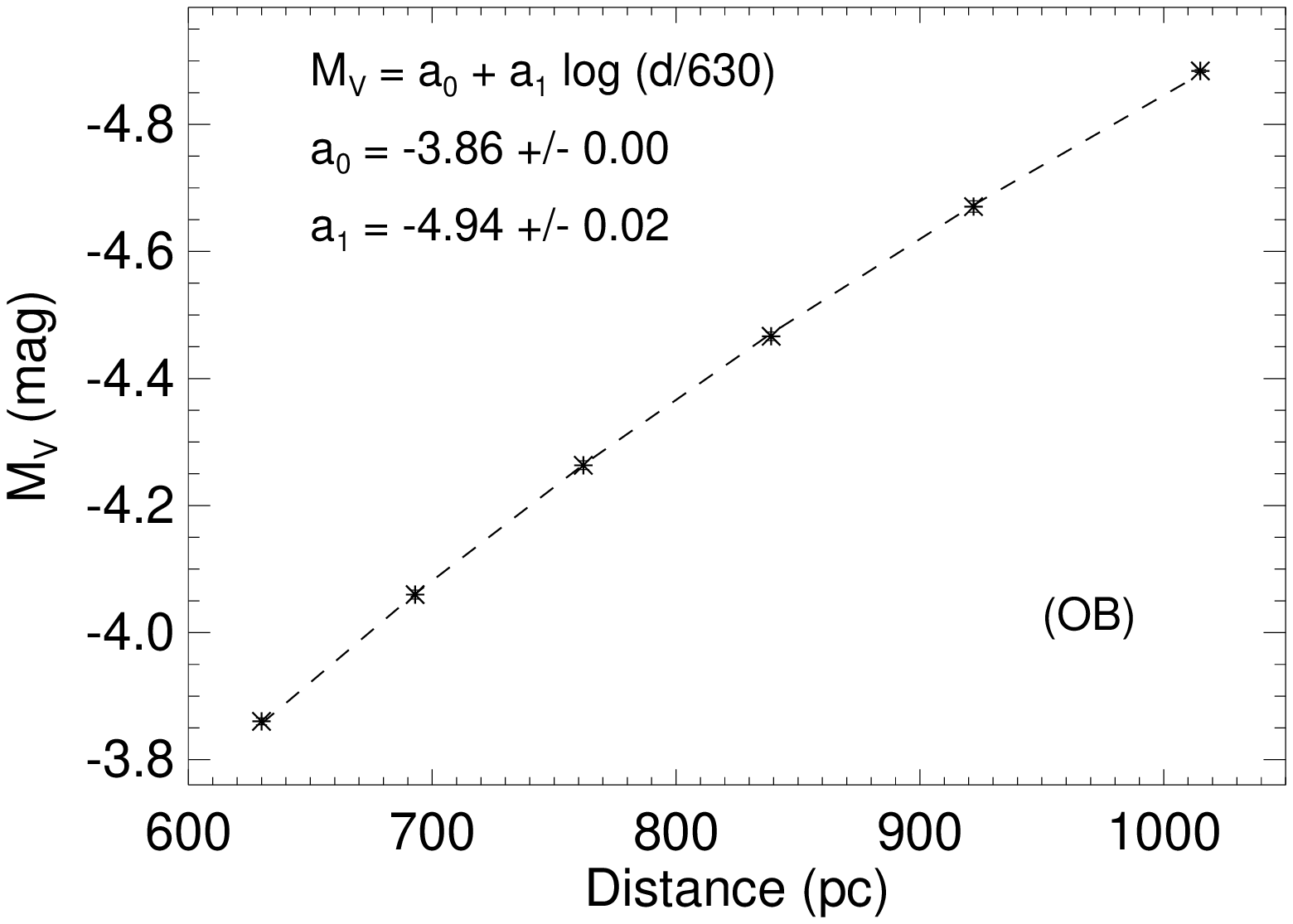}
\end{center}
\caption{Some best-fit results from the grid-modelling of the \GTC
spectra of \WRE. The dependence of the mass-loss rate, stellar
luminosity and absolute visual magnitude on the adopted distance to
\WR is shown in the first three rows. 
Error bars ($1\sigma$ errors from the fits) are smaller than the plot
symbols denoting them, except for the mass-loss rate and luminosity of
the OB component.
}
\label{fig:gtc_fits}
\end{figure*}

\begin{figure*}
\begin{center}
\includegraphics[width=6in, height=4.3in]{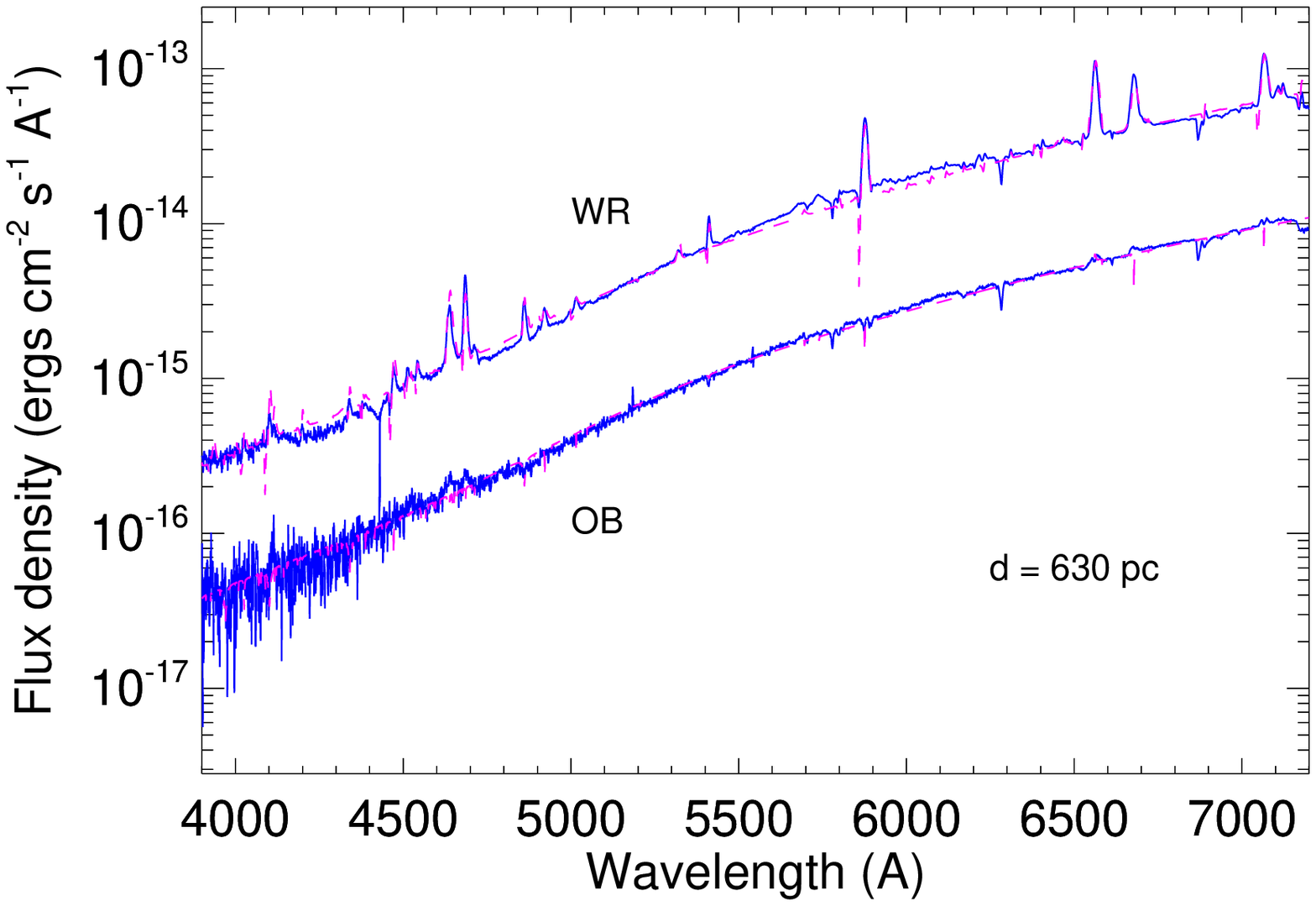}
\includegraphics[width=6in, height=4.3in]{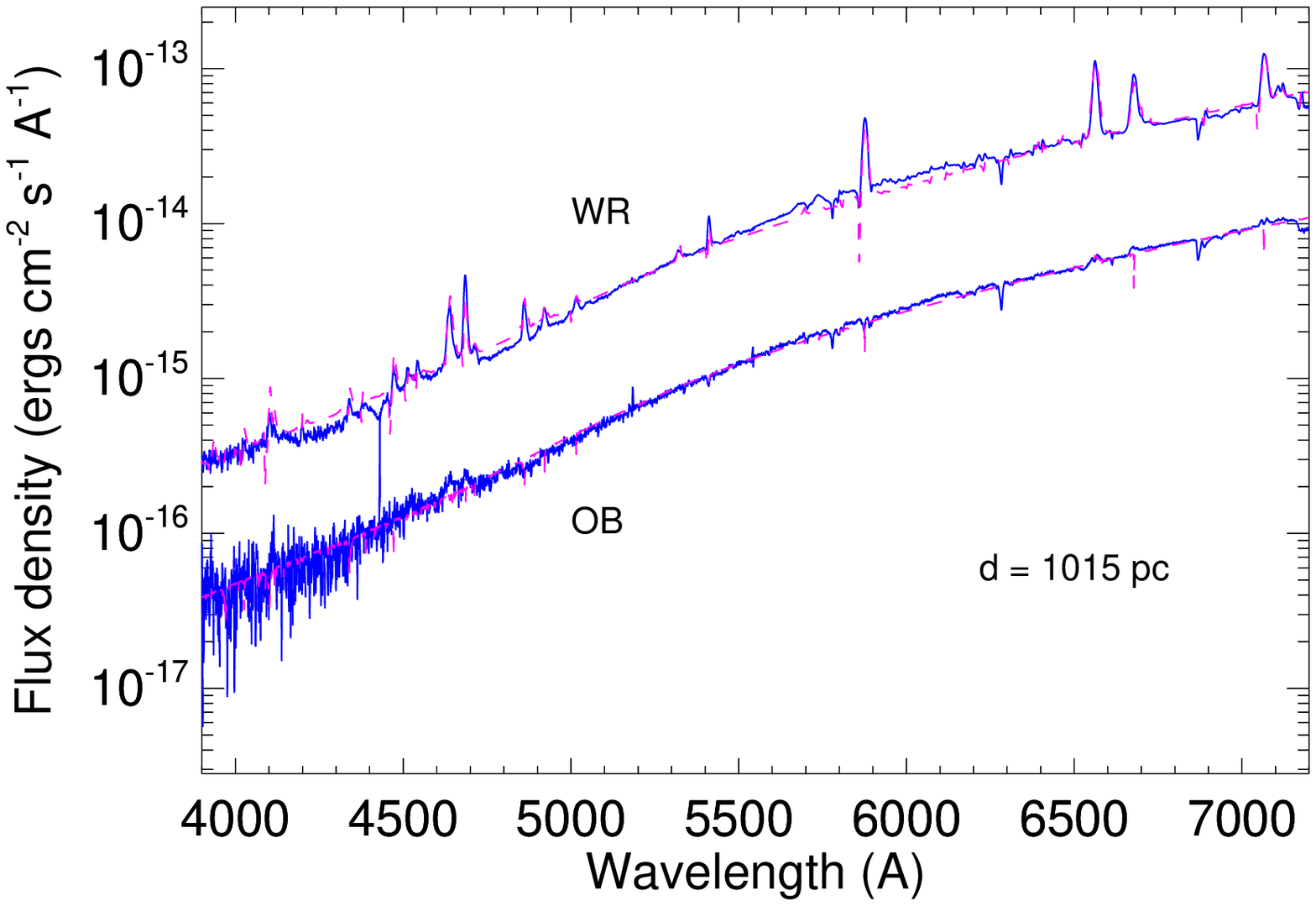}
\end{center}
\caption{Examples of the observed GTC spectra (blue solid line) overlaid 
with the best-fit model (magenta dashed line).
}
\label{fig:gtc_spec}
\end{figure*}

\begin{figure*}
\begin{center}
\includegraphics[width=2.8in, height=2.0in]{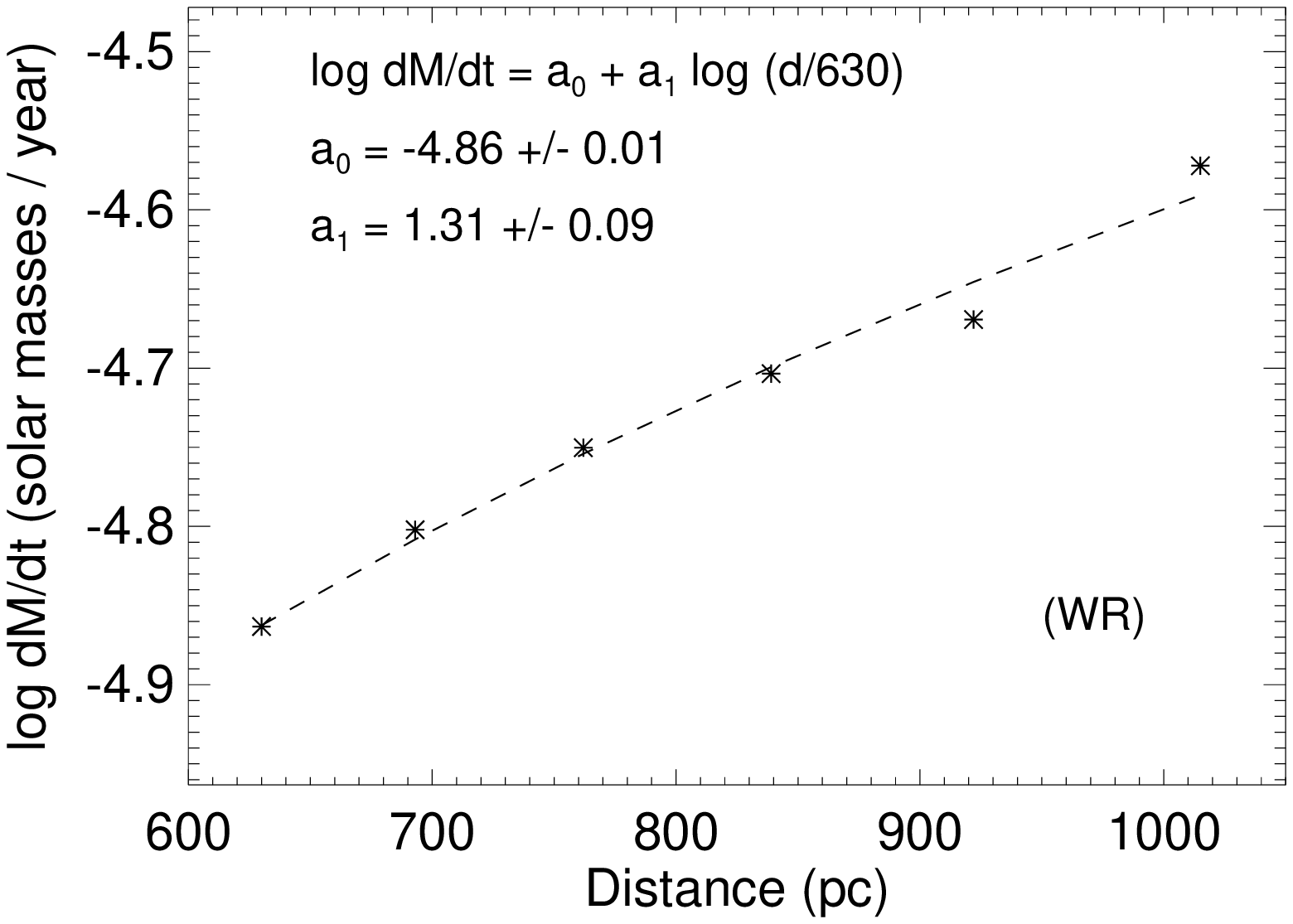}
\includegraphics[width=2.8in, height=2.0in]{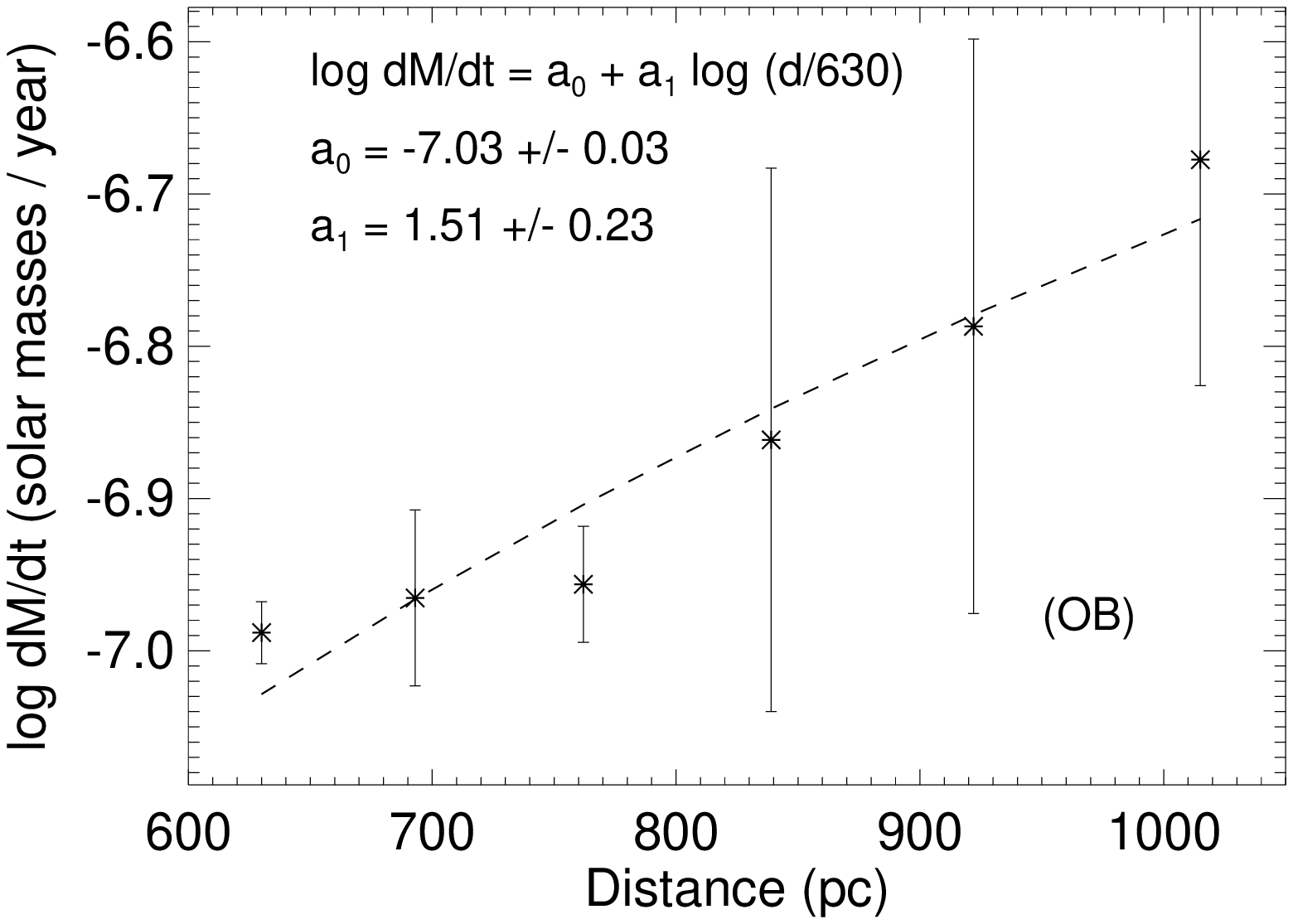}
\includegraphics[width=2.8in, height=2.0in]{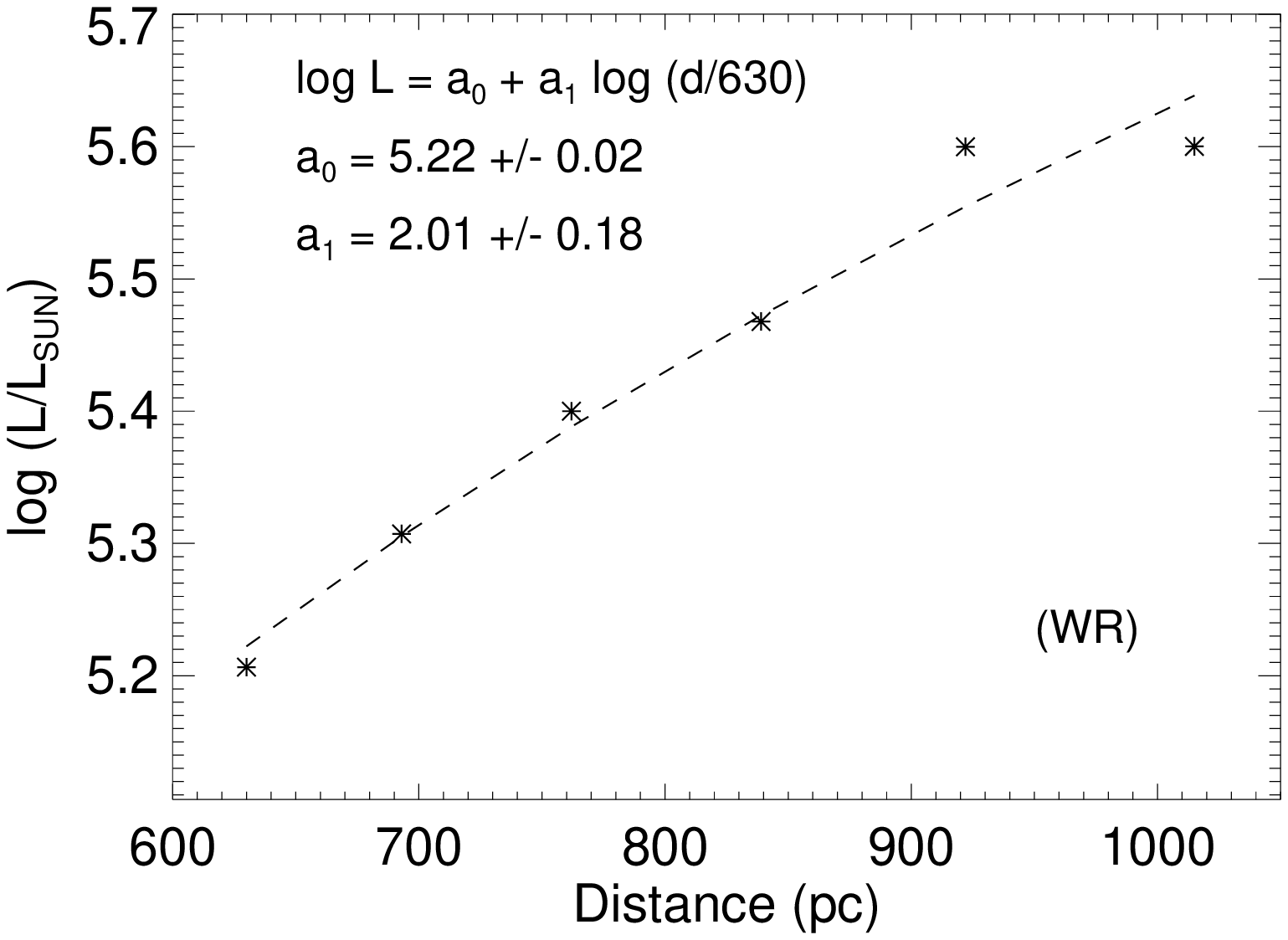}
\includegraphics[width=2.8in, height=2.0in]{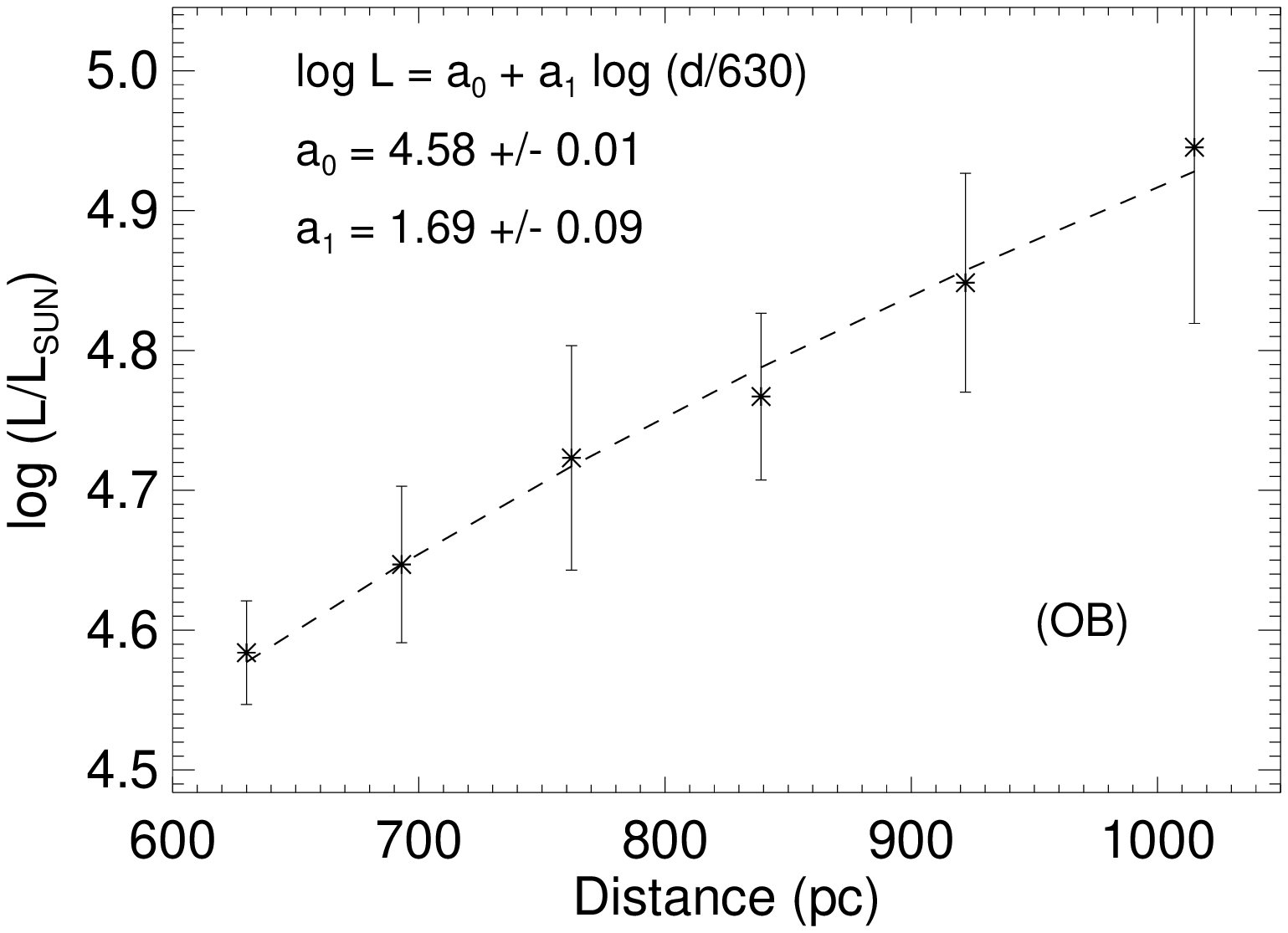}
\includegraphics[width=2.8in, height=2.0in]{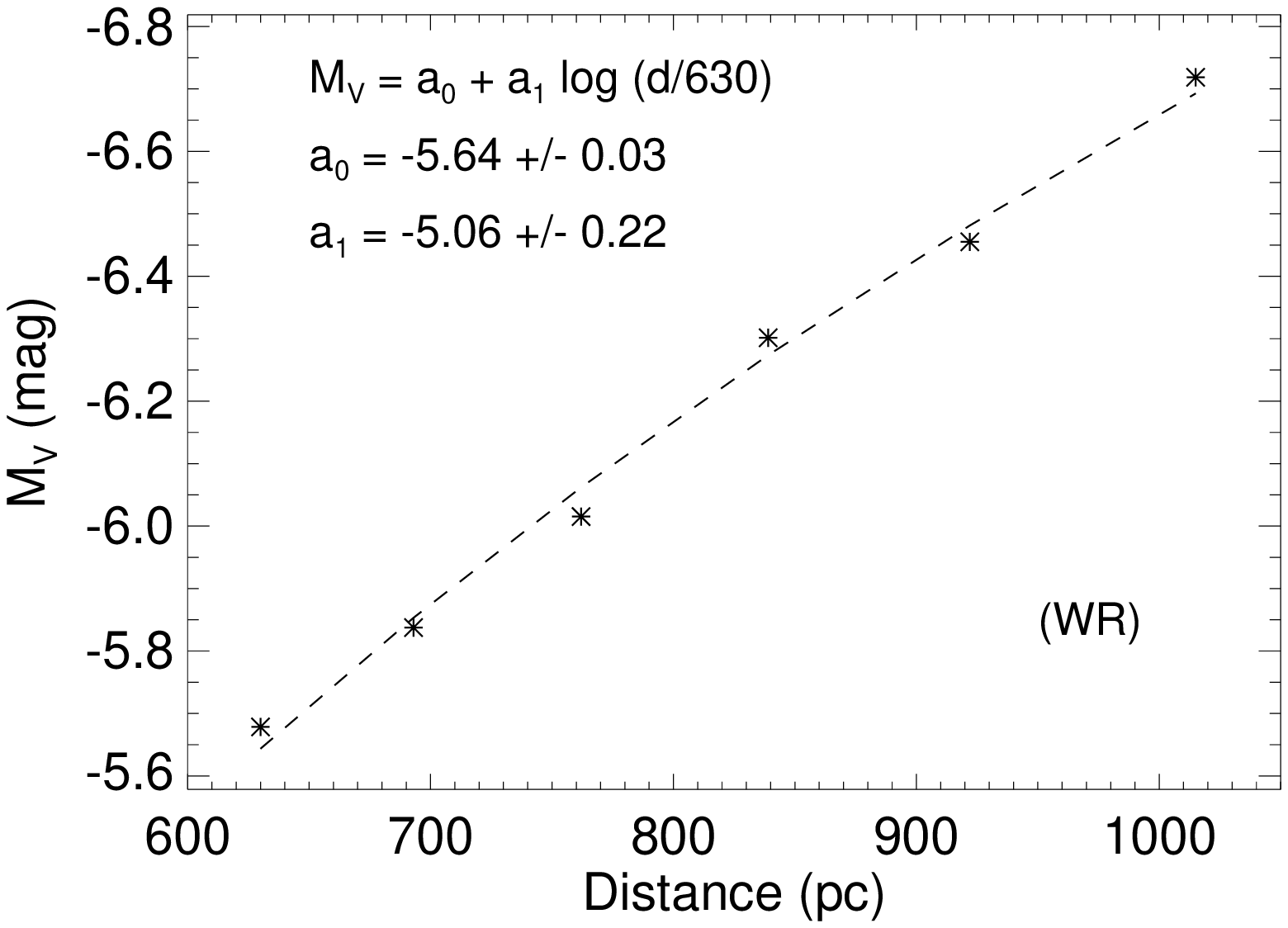}
\includegraphics[width=2.8in, height=2.0in]{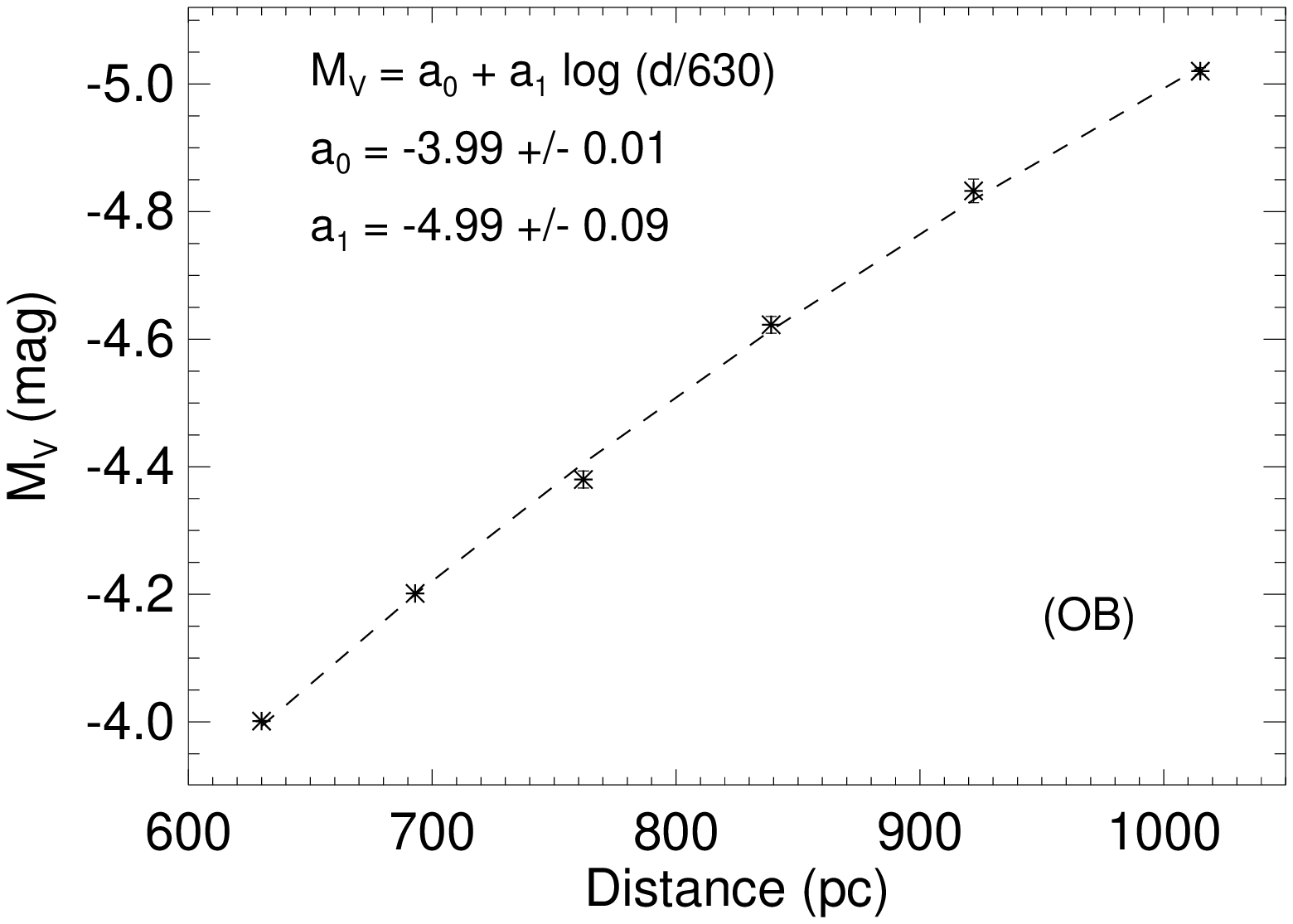}
\end{center}
\caption{The same as in Fig.~\ref{fig:gtc_fits} but for the
grid-modelling of the \HST spectra of \WRE.
}
\label{fig:hst_fits}
\end{figure*}

\begin{figure*}
\begin{center}
\includegraphics[width=6in, height=4.3in]{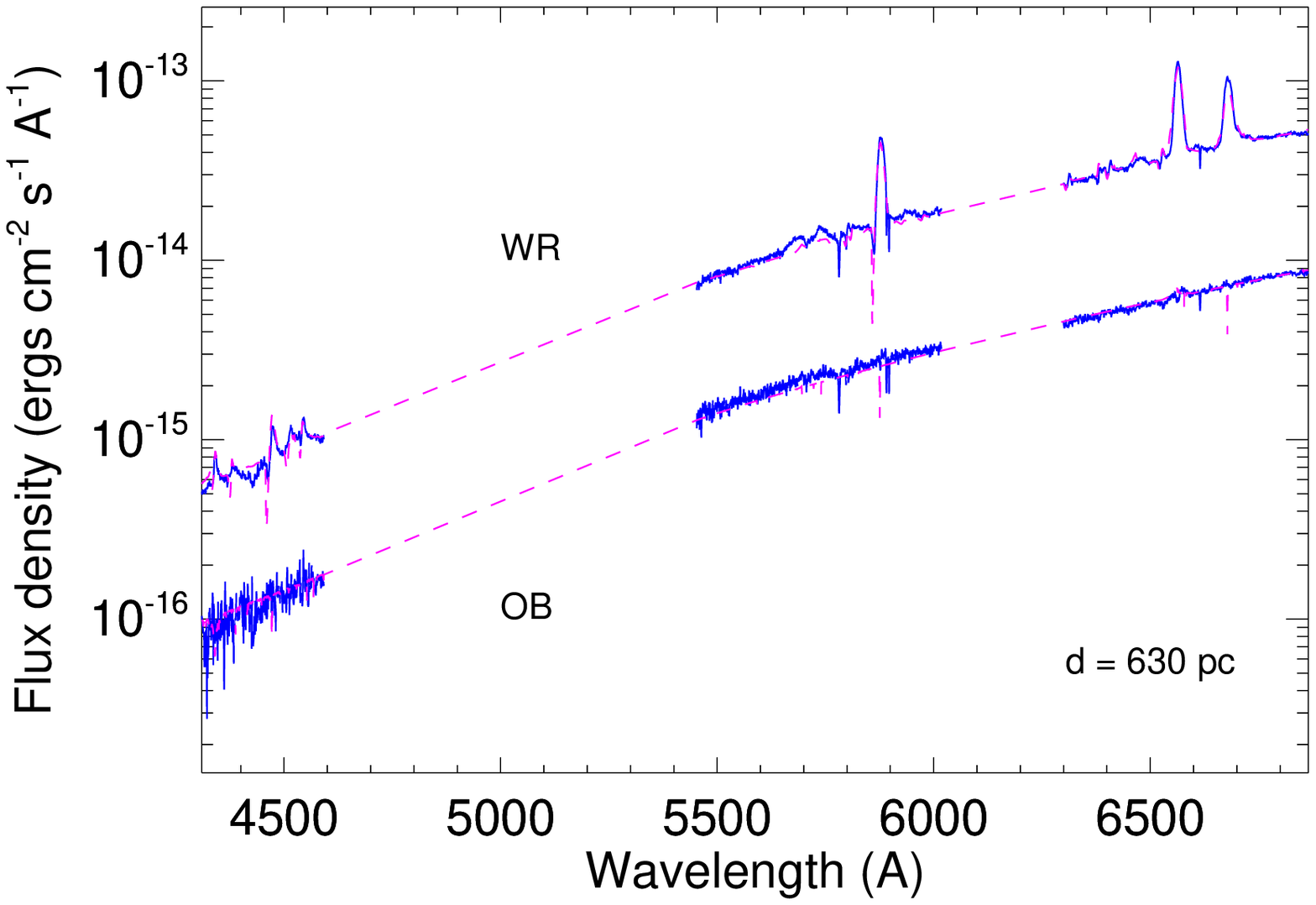}
\includegraphics[width=6in, height=4.3in]{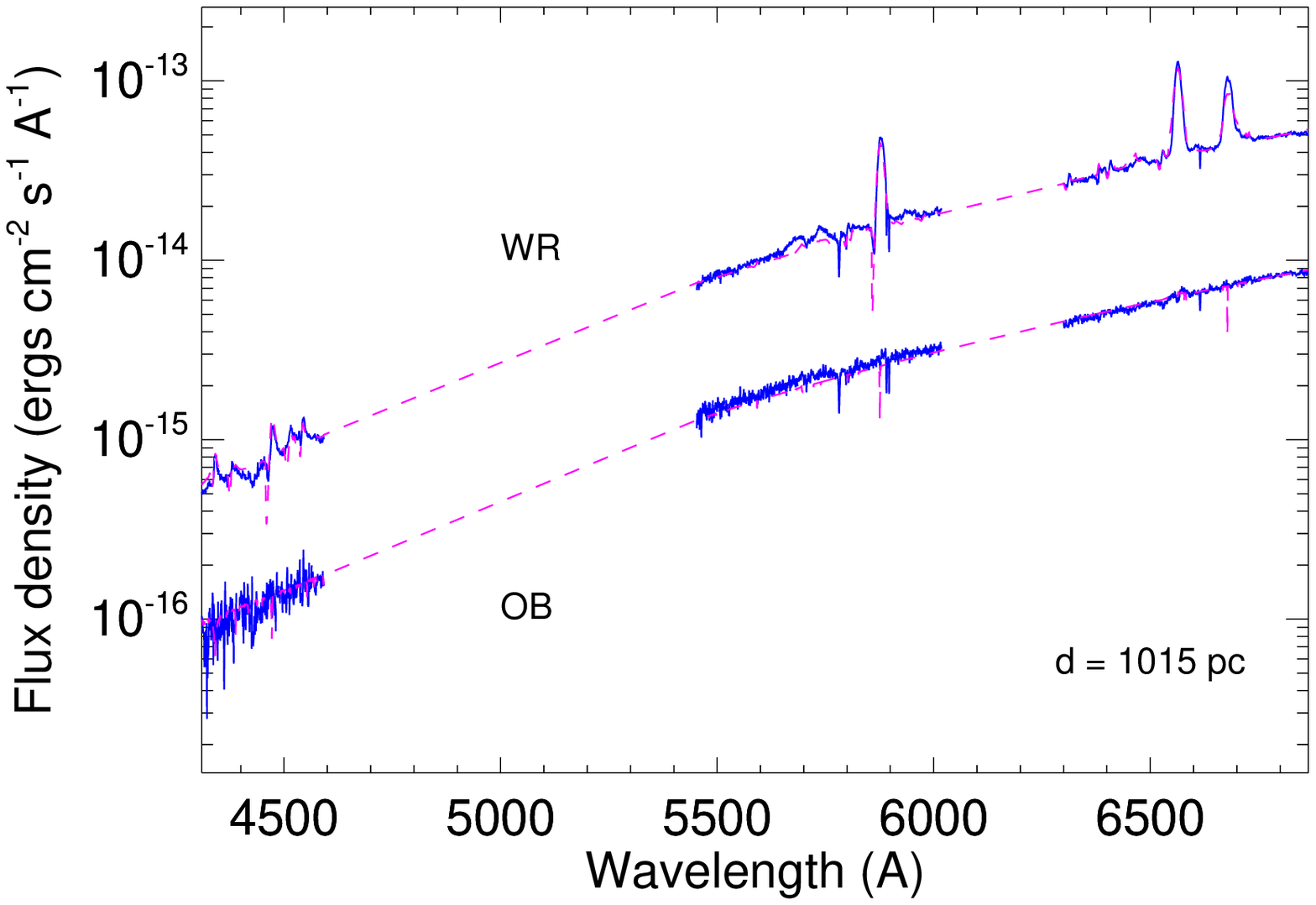}
\end{center}
\caption{Examples of the observed HST spectra (blue solid line)
overlaid with the best-fit model (magenta dashed line).
}
\label{fig:hst_spec}
\end{figure*}

\subsection{Results}
\label{sec:fit_results}
For fitting the observed GTC and HST spectra of \WR
(WN8$+$B0.5 - O9; e.g., \citealt{niemela_98}), we adopted 
a uniform approach for modelling the stellar spectra of both
binary components.  So, we have built
two grids of model spectra that are representative for  WN8 and OB
objects, respectively, and consider the following stellar parameters:
mass-loss rate \dotM (in units of solar masses per year, \dotMyr),
stellar luminosity \logL (in units of solar luminosity, \sunlum) and
effective temperature \Tstar (in Kelvin). The grids are based on 80 
and 60 model spectra for the WN8 and OB star, respectively. 
All models have a spectrum-formation zone with an inner radius
corresponding  to the Rosseland opacity of 100 and an outer radius 
of a hundred inner radii.
The ranges of the basic stellar parameters are as follows. 

For the WN8 grid, we have
\dotM $= [-5.0,~ -4.8,~ -4.5,~ -4.3]$,
\logL $= [5.0,~ 5.2,~ 5.4,~ 5.6,~ 5.8]$,
\Tstar $ = [32500,~ 35000,~ 37500,~ 40000]$. The adopted chemical
abundances of H and He are those from \citet{morris_00} and for all
other elements are from \citet{vdh_86}. The stellar wind velocity is
1000\kms
corresponding to the range of wind velocity values $950 - 1100$\kms 
derived from observations of \WR 
(\citealt{eenens_94}; \citealt{hamann_95}; \citealt{morris_00}).

For the OB grid, we have
\dotM $= [-7.0,~ -6.8,~ -6.3]$,
\logL $= [4,2,~ 4.5,~ 4.8,~ 5.0,~ 5.5]$,
\Tstar $ = [27500,~ 30000,~ 32500,~35000]$, the abundances are
solar \citep{asplund_09} and the stellar wind velocity is 1600\kms
representative of the range of wind velocity values typical for a 
B0.5 - O9 massive star (e.g., \citealt{prinja_90}; 
\citealt{prinja_98}).

The currently accepted distance to \WR is $630\pm70$~pc
\citep{church_92} but it is a photometric distance, so, we explore a
range of values for this parameter. Namely, we consider six different
values starting with d $= 630$~pc and each consecutive value
increases by 10\%. Thus, we could explore how the stellar parameters
derived from the model fitting depend on the adopted distance to \WRE.
It is worth noting that the \Gaia parallax of \WR is negative
($ -1.2733\pm0.3144$; \citealt{gaia_dr2}) 
likely due to the binary nature of this object.
As a result, the \Gaia distance to this object is not well 
constrained yet: it is much larger than the currently accepted 
distance to \WR  and with large uncertainties 
(d $= 4168^{+1619}_{-1174}$~pc; \citealt{bailer_jones_18}).

Based on each best-fit spectrum, we also calculated some synthetic
magnitudes: the Johnson B and V, and the corresponding intrinsic
colour $(B-V)_0$ and absolute visual magnitude M$_V$. For these, we
made use of the filter response functions from \citet{bessell_12} and
zero points from \citet{casagrande_14}.

Anticipating the results from the grid-modelling, we note that
exploring the entire OB grid in a standard way (i.e., fitting for all
grid parameters, \dotME, \logLE, \TstarE)  did not provide well
constrained fit results for the OB star, due to the lack of strong 
spectral feature in its spectrum. So, to obtain some constraints on 
its physical parameters, we ran three different series of models each 
having a fixed value of the stellar temperature; namely, 
T$_* = 30.0,~32.5,~ 35.0$~kK.

Results from the grid-modelling of the \GTC and \HST spectra of
\WR are presented in Figures~\ref{fig:gtc_fits},  \ref{fig:hst_fits}
and Table~\ref{tab:fits}. The following things are worth 
mentioning.

We see that the derived stellar parameters in general follow the
theoretically expected (approximate) correlations with the 
distance to the studied 
object: e.g., $\dot{M} \propto d^{1.5}$, $L \propto d^2$ (see eqs. 1 
and 4 in \citealt{hillier_99}; also \citealt{schmutz_89}) and the 
absolute visual magnitude has its standard dependence on distance:
M$_V \propto 5 \log d$.

The values of the stellar temperature of the WR component are very
similar for the fits to the \GTC and \HST spectra, thus, indicating a
stellar temperature of the WN8 star in \WR of T$_* \sim 37$ kK. And, 
the same is valid for the OB star in this binary, T$_* \sim 32.5$ kK.

Also, the values for the interstellar extinction, derived from the 
fits to the spectra of the WR component in \WR 
(Table~\ref{tab:fits}),  A$_V = 3.1 E(B-V) = 10.42 - 10.54$ mag,
correspond well to that adopted for this object in the VII-th 
Catalogue of Galactic Wolf-Rayet stars  (A$_v = 1.11$ A$_V$ and
table 28 in \citealt{vdh_01}): namely  A$_v = 11.57 - 11.70$ mag
vs. A$_v = 11.60$ mag. On the other hand, the fits to the spectra of 
the OB component in \WR give a slightly higher value of the 
interstellar extinction (A$_v = 12.04$ mag). But, we have to
keep in mind that the quality of the OB-star spectra is not very high,
especially in the blue region, and this could be the reason for such a
difference.

It is worth noting that the interstellar extinction towards \WR might
not be standard ($R_V \neq 3.1$). For example, based on analysis of 
optical and infrared data of \WR  \citet{morris_00} suggest a value of 
$R_V = 2.7$.
However, the wavelength range of the optical spectra in our study is 
not very large ($\sim 3000$~\AA) which prevents constraining the 
$R_V$-value very well. We nevertheless explored the case of non-standard 
interstellar extinction with $R_V = 2.7$ but it provided a 
poorer match to the observed spectra. We thus feel confident in
adopting the standard interstellar extinction curve ($R_V = 3.1$) in 
the current spectral modelling.

The \GTC and \HST synthetic magnitudes are in general consistent between
each other. We also note that the synthetic V magnitudes are very
similar to the observed ones reported by \citet{niemela_98} from \HST
photometry. However, the synthetic B magnitudes are by $\sim 1$ mag
brighter than those reported by these authors, which is likely due to
the different response function of the filters.

On the other hand, the values of synthetic magnitudes in the Johnson V
filter match well those observed 
(see last two rows in Table~\ref{tab:fits}).
The observed V magnitude values of the \WR system are from the ASAS-SN 
(All Sky Automated Survey for Supernovae) campaign for the period 2017 
September 12 - 27\footnote{For ASAS-SN see
\url{http://www.astronomy.ohio-state.edu/~assassin/index.shtml/}} 
(e.g., \citealt{asas_sn_1}; \citealt{asas_sn_2}), thus, bracketing 
the date of the \GTC observation (2017 September 18):
V(ASAS-SN) $= 13.76\pm0.03$~mag.

To broaden such a comparison, we calculated synthetic magnitudes
for the Pan-STARRS survey (Panoramic Survey Telescope and Rapid
Response System)  following the definition of its photometric
system and using the filter bandpasses from \citet{pan_starrs_1}.
The observed magnutides are from the first data release
\citep{pan_starrs_dr1}.
Also, we calculated synthetic \Gaia magnitudes adopting the filter
bassbands and zero points from \citet{gaia_dr2_phot}. The observed
magnutides are from the \Gaia Data Release 2 (DR2;
\citealt{gaia_dr2}).

\begin{figure}
\begin{center}
\includegraphics[width=2.8in, height=2.0in]{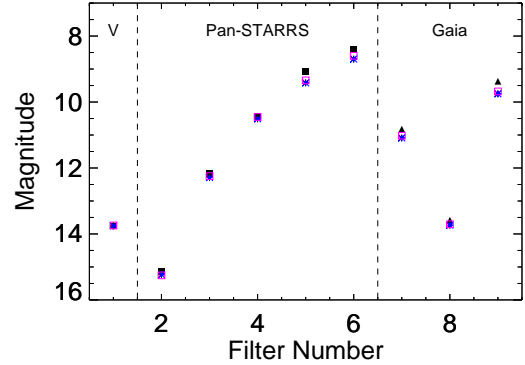}
\end{center}
\caption{
Comparison between synthetic and observed magnitudes of \WRE.
The observed magnitudes are given with filled symbols: a circle for
the V magnitude from the ASAS-SN campaign (filter number 1, 
$\lambda_{mean} =  5500$~\AA;
$\lambda_{mean} =  \sum \lambda T_{\lambda} / \sum T_{\lambda} $, 
$T_{\lambda}$ is the filter passband at wavelength $\lambda$); 
squares for the g, r, i, z , y 
magnitudes from the Pan-STARRS catalogue (filter numbers 2 - 6, 
$\lambda_{mean} =  4866, 6215, 7545, 8680, 9633$~\AA, 
respectively); triangles for the g, bp, rp magnitudes from the \Gaia 
DR2 catalogue (filter numbers 7 - 9, $\lambda_{mean} =  6406, 
5131, 7777$~\AA, respectively).
The synthetic magnitudes for the best GTC and HST models are marked
with asterisks (in blue) and open squares (in magenta), respectively.
\label{fig:all_mag}
}
\end{figure}

As seen from Fig.~\ref{fig:all_mag} and 
keeping in mind that our synthetic magnitudes are based on modelling
two different spectral sets (GTC and HST ) as each spectral and
photometric data set bears its own observational and systematic
uncertainties, we think that the correspondence between the 
synthetic and observed magnitudes is at acceptable level.

So, all this gives us confidence in the results from the performed
grid-modelling of the optical spectra of \WRE.
We think that the stellar parameters of the WR component in \WR are 
acceptably well constrained while those of the OB component are not
so well constrained. The latter is understandable given the quality 
of the spectra (both \GTC and \HSTE) of the OB star in \WRE.
Nevertheless, we note that for the best-fit values of the intrinsic
(B-V)$_0$ colour and the absolute visual magnitude M$_V$ are typical
for an OB object of B0.5 - O8 V spectral type (e.g.,
\citealt{kaler_82}): corresponding to the range of distances 
explored here.

We would also like to mention
that modelling of the stellar spectra could in general provide
information on the chemical composition of the emitting gas. However, 
there are no resonance lines in the optical spectra of \WRE, which 
does not allow deriving very accurate values for the chemical 
abundances. We thus only tried to derive an estimate of the nitrogen 
abundance: it is the most abundant light metal in the WN stars. 
We recall that \citet{morris_00} derived a hydrogen-to-helium 
abundance (by number) of 0.4, adopted in our study.
For the best-fit case from the grid modelling of the WN component 
in \WRE, we used the \cmfgen code to get a theoretical spectrum only
of nitrogen. By varying its contribution to the total spectrum, 
the fit of the observed spectrum gave an estimate of the 
nitrogen-to-helium abundance  (by number): 
N / He $= 4.4\pm0.2\times10^{-3}$.

Finally, we assume that the derived values of the stellar parameters 
from the fits to the \GTC and \HST spectra of \WR could in fact 
define a possible range of values (accuracy) for 
a given stellar parameter, assuming a fixed/known distance.  
Thus, although a given physical parameter may vary appreciably for 
the explored range of distances to \WR (see above), 
we note that for each distance value 
the accuracy of mass-loss rate of both stellar
components is $\sim 0.1$ dex, that of the stellar luminosity is 
$\leq 0.1$ dex (WR) and $\sim 0.1 - 0.15$ dex (OB), and that of the
absolute V magnitude is $\sim 0.15$ mag for both components.

Because the stellar parameters are key ingredients in the general
physical picture of \WRE, specifically the mass-loss rate (see
Section~\ref{sec:discussion}), it is interesting to compare the
values derived in this study with those from previous analyses using
stellar atmosphere models by \citet{morris_00} and 
\citet{hamann_19}\footnote{Note that the model results in 
\citet{hamann_19} are in fact those of \citet{hamann_06}, since no 
adjustment of the distance value to \WR was possible based on the 
\Gaia DR2 results.}.
For the distances adopted in those studies (650 pc in
\citealt{morris_00}; 1200 pc in \citealt{hamann_19}, this value
corresponds to their adopted value for the distance modulus of 10.4
mag) and the same volume filling factor ($f = 0.1$), 
the mass-loss rate of \citet{morris_00} is about 2 times higher and
 that from \citet{hamann_19} is about 3 - 4 times higher than the 
value derived here.
We think that one of the reasons for such differences is related to 
the fact that the absolute calibration of the models in the latter is
based on adopting some absolute magnitude typical for the object 
(e.g., absolute $K$-magnitude in \citealt{morris_00}; absolute
$\varv$-magnitude in \citealt{hamann_19}), while our grid modelling is
based on fitting directly for the stellar luminosity.  It is our
understanding that fitting for luminosity should be physically
reasonable. On the other hand, absolute magnitudes of Wolf-Rayet stars
even of a given subclass may have an appreciable scatter (e.g.,
\citealt{vdh_01}; \citealt{wr_absmag_20}).
A limitation of our analysis is that it is based only on the optical
spectra of \WRE, while both other studies take into consideration
optical-infrared data. However, the good correspondence between the
results of our modelling and observations of a range of facilities
(e.g., see Fig.~\ref{fig:all_mag}) gives us additional confidence in 
the derived results.
But, we note that a factor of $\sim 2$ difference could in general
reflect uncertainties typical for deriving mass-loss rates based on
using stellar atmosphere models (e.g., based on different atomic data
etc.).

\begin{table}
\caption{\WR Spectral Fit Results
\label{tab:fits}}
\begin{tabular}{lllll}
\hline
\multicolumn{1}{c}{Para-} & 
\multicolumn{2}{c}{WR }  & \multicolumn{2}{c}{OB }  \\
\multicolumn{1}{c}{meter} &
\multicolumn{1}{c}{GTC}  & \multicolumn{1}{c}{HST} &
\multicolumn{1}{c}{GTC}  & \multicolumn{1}{c}{HST} \\
\hline
\Tstar  & $36.9\pm0.9$  & $37.3\pm0.8$   &  $30.0 - 35.0$  &
$30.0 - 35.0$ \\
E(B-V)   & $3.36\pm0.01$ & $3.40\pm0.01$  & $3.50\pm0.01$  &
$3.50\pm0.01$  \\
(B-V)$_0$   & $-0.29\pm0.01$  & $-0.29\pm0.01$  & $-0.30\pm0.01$  &
$-0.30\pm0.01$  \\
B   & $16.86\pm0.01$  & $16.87\pm0.01$  & $19.02\pm0.01$  &
$18.89\pm0.01$  \\
V   & $13.92\pm0.01$  & $13.89\pm0.01$  & $15.99\pm0.01$  &
$15.86\pm0.01$  \\
V$_{sys}$   & \multicolumn{2}{c}{$13.77\pm0.01$ (GTC)} &
\multicolumn{2}{c}{$13.72\pm0.01$ (HST)}  \\
            &  \multicolumn{4}{c}{$13.76\pm0.03$ (ASAS-SN)}  \\
\hline

\end{tabular}

{\it Note}.
Results from the fits to the optical spectra of \WRE.
Tabulated parameters are the stellar temperature \TstarE, interstellar
extinction E(B-V), intrinsic colour (B-V)$_0$, 
B, V and total magnitude (WR$+$OB) of the system V$_{sys}$
followed by the observed V magnitude from the ASAS-SN campaign.
The value of \Tstar is in unit of kK ($10^3$ Kelvin) and all other 
parameters are in magnitudes.
The errors are the standard deviation for the fit results of a given 
parameter as derived for the sample of six values of the distance to 
\WR (see text).

\end{table}

\section{Discussion}
\label{sec:discussion}
\subsection{CSW picture}
\label{sec:csw}
Global fits to the GTC and HST spectra of \WR allowed us to derive
some basic physical parameters of both stellar components of this
WR$+$OB system (see Section~\ref{sec:fit_results}), which
could be used to check the general physical picture
for this colliding-wind binary. 

We recall that mass-loss rate is a key parameter for modelling the 
X-ray emission from CSW binaries 
(\citealt{luo_90}; \citealt{stevens_92}; 
\citealt{mzh_93}). Analysis of the \Chandra data on \WRE, based on 
hydrodynamic simulations, showed that the mass-loss rate of the WN8
component in \WR should be $\sim 10^{-5}$\dotMyr (for adopted distance
of 630 pc) in order to match the observed X-ray flux from the CSW
region in \WR \citep{zhp_10b}. 
We see that the values of mass-loss rate derived from modelling the
optical spectra of \WR are in acceptable correspondence with that from
modelling its X-ray emission: \dotM $= -4.95; -4.86$~(GTC; HST; see
Figs.~\ref{fig:gtc_fits},~\ref{fig:hst_fits}) vs.  
\dotM $= - 5.0$ (\ChandraE).

It is worth noting that the results from the grid-modelling of the
optical spectra of both stellar component in \WR (Section~
\ref{sec:fit_results}; Figs.~\ref{fig:gtc_fits}, ~\ref{fig:hst_fits})
suggest that the wind momentum ratio is 
$\dot{M}_{WR} V_{WR} / \dot{M}_{OB} V_{OB} = 67 - 54$ for the range of
distance to this object considered here. These values are larger than
the one deduced from high-resolution optical and radio observations of
$\sim 36$ (e.g., $\dot{M}_{OB} V_{OB} / \dot{M}_{WR} V_{WR} = 0.028$;
\citealt{niemela_98}).
A reason for such a mismatch is that the fits to the optical spectra
of the OB component suggest low mass-loss rate since no appreciable
emission lines are present in its spectrum. However, if the stellar
wind has larger velocity than the adopted in the grid-modelling, this
might be a solution to the wind-momentum-ratio mismatch.

To explore this, we constructed a grid of models for the OB star that
has stellar temperature of \Tstar $= 32.5$~kK, stellar luminosity and
mass-loss rate as in the original grid and wind velocity 
$V_{OB} = [1600, 2000, 2500]$\kmsE.
Once we have derived the best-fit parameters for the WR component, we
know its wind-momentum $\dot{M}_{WR} V_{WR}$. Then, the grid-modelling
of the OB spectra {\it requires} the wind-momentum ratio to be
$\dot{M}_{WR} V_{WR} / \dot{M}_{OB} V_{OB} = 36$, which means that the
$\dot{M}_{OB}$ and $V_{OB}$ parameters are {\it not} independent. 
Namely, the $V_{OB}$ parameter is fitted for while the $\dot{M}_{OB}$
parameter gets the corresponding value that satisfies the {\it
required} wind-momentum ratio of 36.

Such a grid-modelling provided fits to the optical spectra of the OB
component in \WR with the same quality as from the original grid. The
values of the mass-loss rate were similar as well and the stellar wind
velocity was in the range 1900 - 2300\kmsE. 
We note that such terminal wind velocities are not atypical for a B0.5
- O8 V massive star (e.g., \citealt{prinja_90}; \citealt{prinja_98}).

We thus see that results from analysis of optical and X-ray properties
of \WR do `converge' in the framework of the CSW picture of massive
binaries.

\subsection{Radio Emission}
\label{sec:radio}
Wolf-Rayet stars are radio sources (e.g.,
\citealt{abbott_86}; \citealt{leitherer_97} and references therein) and
the level of their thermal radio emission depends on the mass-loss rate
(\citealt{pa_fe_75}; \citealt{wri_bar_75}): $S_{\nu} \propto
\dot{M}^{4/3}$, where $S_{\nu}$ is the flux density and $\dot{M}$ is
the mass-loss rate. In the case of stellar winds with clumping and no
emission from the inter-clump space (as assumed in the stellar
atmosphere models), \citet{abbott_81} showed that  
$S_{\nu} \propto \dot{M}^{4/3} / f^{2/3}$, 
where $f$ is the volume filling factor.
So, radio observations can be used to better constrain the stellar
parameters derived from analysis of optical spectra of massive stars.

We recall that \WR was spatially resolved in high-resolution radio 
observations: its southern component (the WN8 component in the
binary), is a thermal source, while its northern counterpart (located 
between the WN8 and OB components in the binary), is a non-thermal 
source (\citealt{moran_89}; \citealt{church_92}; 
\citealt{co_etal_96}; \citealt{williams_97}; \citealt{co_ro_99};
\citealt{skinner_99}).
For technical consistency, we made use of the published radio fluxes
of the WN8 component in \WRE, obtained {\it only} with VLA (Very Large 
Array; National Radio Astronomy Observatory, USA), to construct its 
radio spectrum. Then, we compared the theoretical radio spectrum of 
the WN8 component in \WRE, based on the derived best-fit stellar 
parameters from the grid-modelling (see Section~\ref{sec:fit_results}), 
with that observed. 

First, we calculated theoretical spectrum in the framework of the
standard radio model by adopting eqs. (1 - 4) from
\citet{leitherer_97} and taking into account the stellar wind
clumping (i.e., substituting $\dot{M} \rightarrow \dot{M} / f^{1/2}$, 
see above, also eq. 9 in \citealt{hillier_99}):
\begin{equation}
  S_{\nu} = 2.32\times10^4 
            \left(\frac{\dot{M} z}{\varv_{\infty} \mu}\right)^{4/3}
            \left(\frac{\gamma g_{\nu} \nu}{f d^3}\right)^{2/3}
   \label{eqn:radio}
\end{equation}
where $S_{\nu}$ is the flux density in mJy, 
$\dot{M}$ is the mass-loss rate, $\varv_{\infty} $ is the stellar wind
velocity,
$\mu$ is the mean molecular weight, $z$ is the rms ionic
charge, $\gamma$ is the mean number of electrons per ion, $g_{\nu}$ is
the free-free Gaunt factor at frequency $\nu$, $d$ is the distance to
the studied object in kpc and $f$ is the volume filling factor.

The mean molecular weight ($\mu = 3.21$), the rms ionic charge ($z =
1$)  and the mean number of electrons per ion ($\gamma = 1$) had
their values as derived from \cmfgenE, while the mean stellar wind 
temperature was T$ = 9400$ K \citep{church_92}.
We recall that the volume filling factor in our grid models was $f =
0.1$ and for the standard distance of 630 pc to \WR the derived mass-loss 
rate of the WN8 component was 
$\log~\dot{M} = -4.95$ (\GTCE) and $-4.86$ (\HSTE) \dotMyr.

\begin{figure}
\begin{center}
\includegraphics[width=2.8in, height=2.0in]{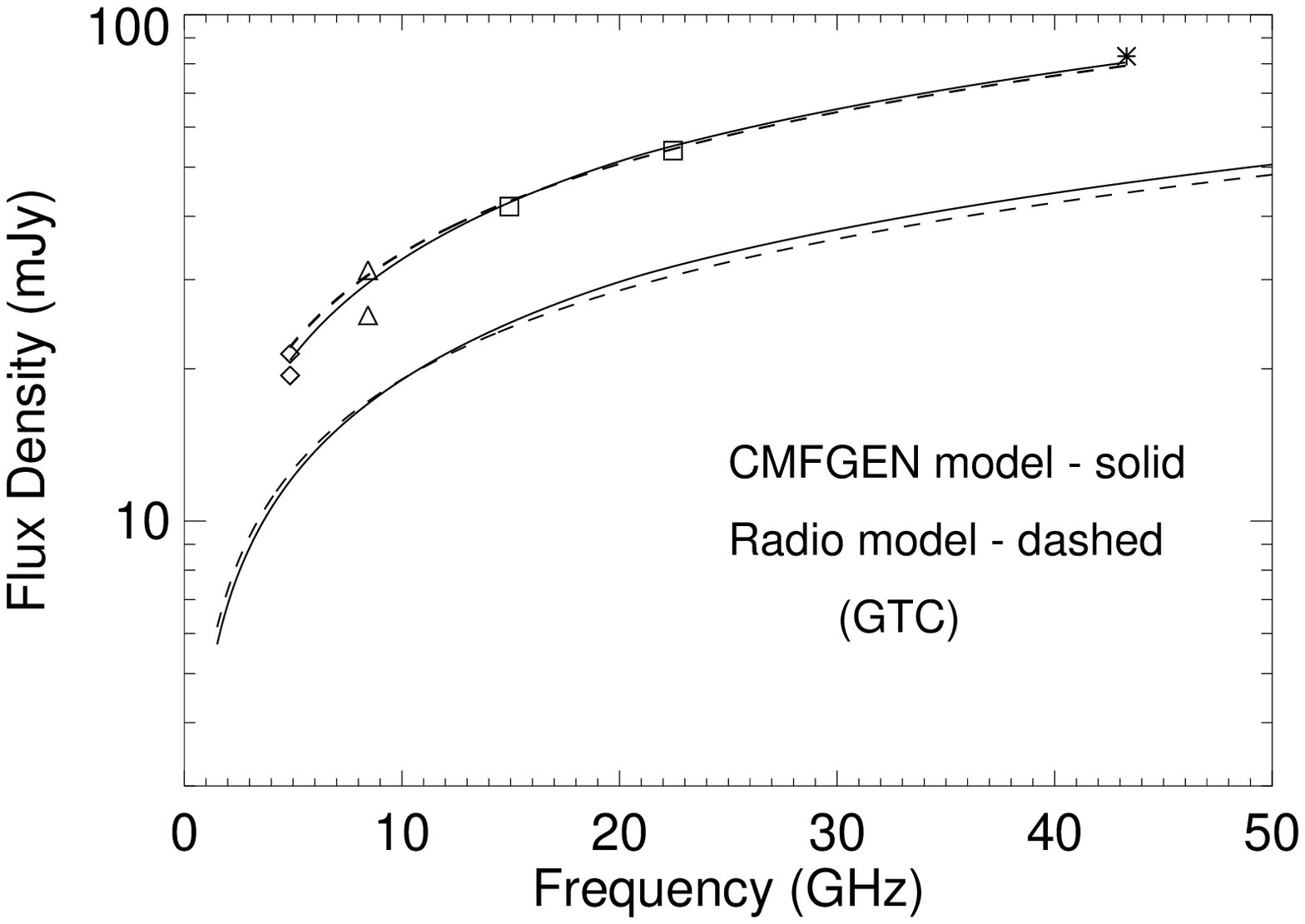}
\includegraphics[width=2.8in, height=2.0in]{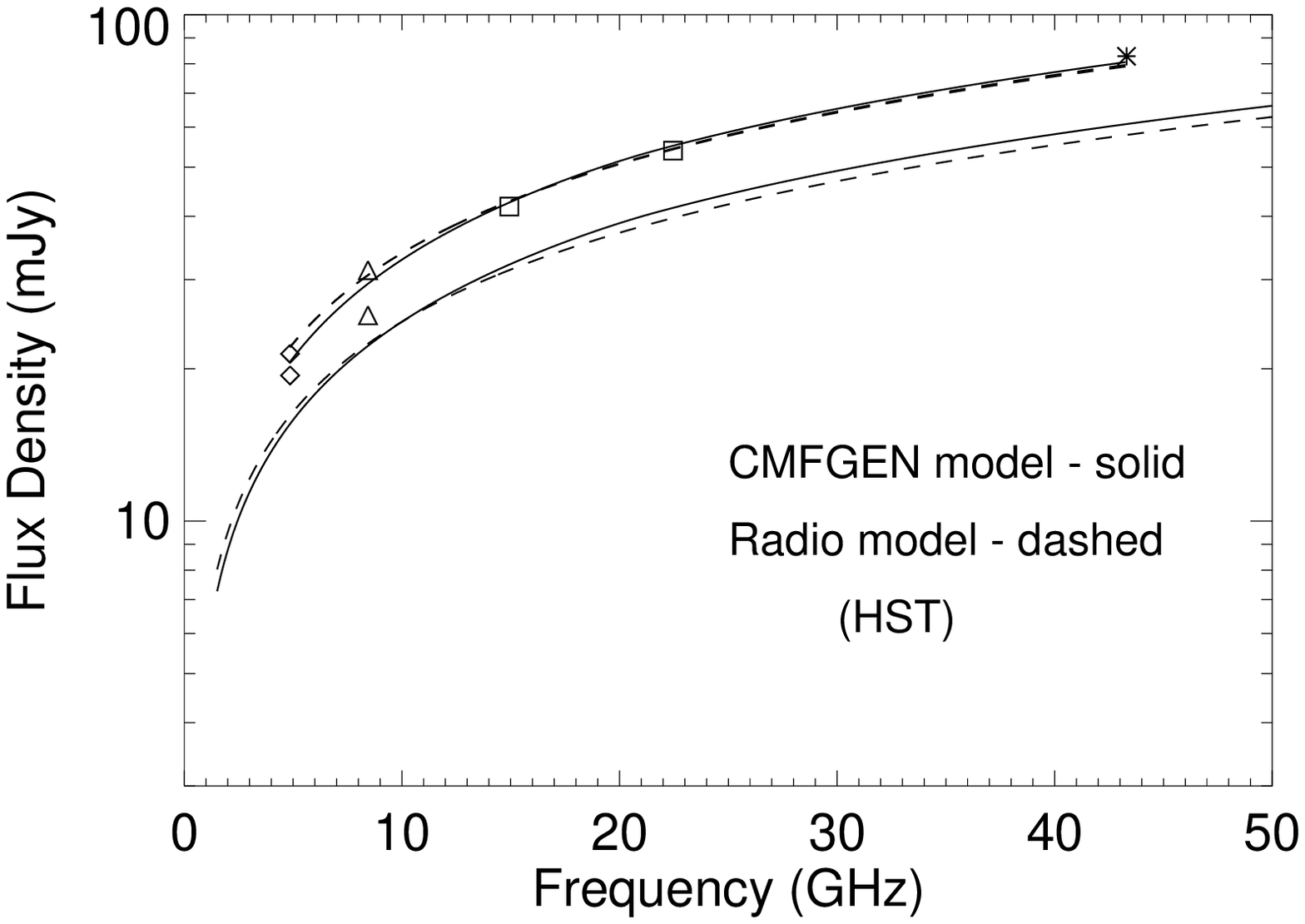}
\end{center}
\caption{{\it Thermal} radio emission of the WN8 component in \WRE.
The observed spectrum is denoted by individual symbols: diamonds
\citep{church_92}, triangles \citep{co_ro_99}, squares
\citep{skinner_99}, asterisk \citep{co_etal_96}. Theoretical spectra
in the 1 - 50 GHz range from \cmfgen and the standard radio model as 
well as their fits to the observed spectrum are shown by lines.
}
\label{fig:radio}
\end{figure}

Second, we calculated theoretical radio spectrum {\it directly} with 
\cmfgen by extending the outer radius of the spectrum-formation zone
to 50 000 stellar radii. 
Figure~\ref{fig:radio} shows the observed radio spectrum of the WN8
component in \WR overlaid with the theoretical spectra for the 
standard distance of 630 pc to this object The following things are
worth noting.

Theoretical radio spectra from \cmfgen and the standard radio model 
match each other acceptably well, which means that either of them can
be used to successfully model radio emission from massive stars.

However, both models predict lower radio flux than observed from the
WN8 component in \WR by a factor of 1.73 - 1.78 (\GTCE) and of 1.33
- 1.37 (\HST): this scaling comes from directly fitting the observed
spectrum by the models (Fig.~\ref{fig:radio}).
To have a good correspondence between the observed radio flux
and the model prediction we need to increase $\dot{M}$ or to 
decrease $f$: $S_{\nu} \propto \dot{M}^{4/3} / f^{2/3}$
(see eq.~\ref{eqn:radio}).

In either case, the observed radio spectrum will be matched well by
the theoretical one, but on the other hand the quality of the model 
fit to the optical spectrum will deteriorate. It is so since the
stellar atmosphere model spectra with clumping taken into account are
sensitive to the $\dot{M}/f^{1/2}$ quantity (e.g.,
\citealt{hillier_99}) and if we change just one of
this parameters, the model spectrum will deviate from its best fit
case.

Alternatively, since the mass-loss rate cannot change with the radius,
could it be that the volume filling factor changes and it becomes
smaller at large radii? This could help increasing the radio flux,
which is produced mostly by the stellar wind, i.e. at large stellar
radii. It is our understanding that such a case, although 
seemingly speculative, might well be physically meaningful.

We recall that in the physical picture adopted in the modern stellar
atmosphere models clumps occupy only a fraction of the volume and
the inter-clump space is `void'. In such a case, clump should expand
with the stellar radius (moving away from the stellar surface) but its
expansion velocity cannot be much larger than the sound speed of the
clump's plasma (see \S 28 in \citealt{zeldovich_raizer_67}). Since 
massive stars have fast stellar winds (from a few hundreds to a few 
thousands \kmsE), expansion of clumps might not be able to `catch up' 
with the expansion of the stellar wind space. As a result,  volume 
filling factor at large radii (in radio-formation zone) might become 
smaller than its value in the UV/optical-formation zone.
A simplified quantitative consideration of this case is given in
Appendix~\ref{app}.

\begin{figure}
\begin{center}
\includegraphics[width=\columnwidth]{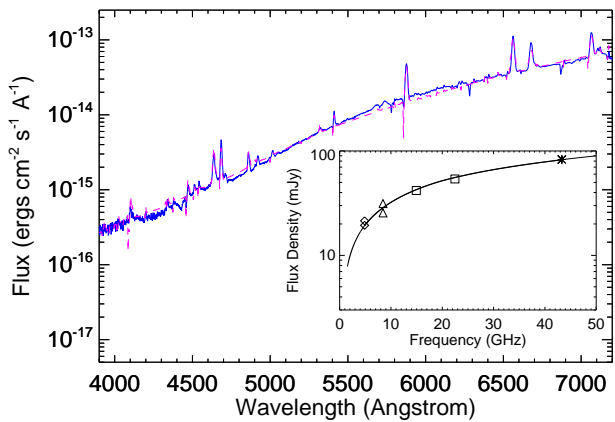}
\end{center}
\caption{The observed \GTC spectrum of the WN8 component in \WR
(blue solid line) overlaid with the \cmfgen model (magenta 
dashed line) using modified volume filling factor
(see Fig.~\ref{fig:clumping}, right panel).
The observed radio emission of the WN8 component (as in
Fig.~\ref{fig:radio}) and the corresponding theoretical spectrum are 
shown in the inset figure.
}
\label{fig:best_clumping}
\end{figure}

An example of spectral results using the modified volume filling
factor is shown in Fig.~\ref{fig:best_clumping}. In this case, we ran
\cmfgen with the best-fit parameters for the \GTC spectrum of \WR and
adopted distance of 630 pc. The fit to the observed optical spectrum 
of the WN8 component in \WR was as good as from the grid-modelling 
giving the same reddening. Also, the theoretical radio emission is 
in very good correspondence with the observed radio spectrum.

\begin{figure}
\begin{center}
\includegraphics[width=2.8in, height=2.0in]{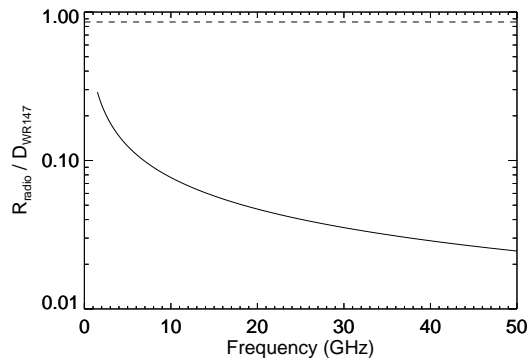}
\end{center}
\caption{The characteristic radius of the radio
emission of the WN8 component in \WR
normalized to the binary separation ($D_{\WRE}$). The dashed line
denotes the location of the colliding-wind zone.
}
\label{fig:r_photo}
\end{figure}

\subsection{Global physical picture}
\label{sec:global_view}
All this (Section~\ref{sec:csw},~\ref{sec:radio}) shows that a global 
view on the colliding-wind binary \WR
could be a reasonable approach for describing its observational 
characteristics in a consistent way. Namely, analysis of optical
spectra with the help of grid-modelling, based on  modern stellar
atmosphere models (e.g., \cmfgenE), provides physical parameters of 
the massive stars which explain well not only the optical spectrum but
the radio and X-ray emission from this colliding-wind binary.
A basic `ingredient' in the physical picture is the presence of
non-homogeneous
stellar winds as the volume filling factor of the clumps should
decrease with distance from the massive star.
And, there is an important consequence from the latter.

In a massive star, UV/optical emission forms relatively close to the
star: at distance smaller than a few tens or a hundred stellar 
radii. Its radio-formation region extends to at least a few thousands
stellar radii. And in wide massive binaries, there is strong X-ray 
emission that originates from the interaction zone of the stellar
winds of the binary components, which could be located very far from
the stars themselves. 
This is indeed the case for the CSW binary \WRE:
the ratio of the characteristic radius of the radio 
emission \footnote{
In the standard radio model \citep{wri_bar_75}, the
radio flux at a given frequency is equal to the integrated flux exterior 
to a radial distance corresponding to a radial optical depth
from infinity of $\tau_{\nu} = 0.244$ 
(eq. 11 in \citealt{wri_bar_75}). 
}
of its WR component to binary 
separation is shown in Fig.~\ref{fig:r_photo}. We see that the X-ray
formation zone, that is the CSW zone, is located far beyond the
radio-formation region. 

Then, since the stellar winds of massive stars
are clumpy (with rather small volume filling factor for the clumps;
e.g. $f \leq 0.1$ in \WRE),
how do clumps manage to collide and form the CSW zone? We recall that
if CSW region does exist, then clumps are quickly heated up and
dissolved in it after having crossed the shock fronts of the CSW zone
(e.g., \citealt{pittard_07}).
It thus seems reasonable to propose that stellar winds of massive
stars are {\it two-component} flows: the more massive component
consists of dense clumps that occupy a small fraction of the stellar
wind volume, and there is a tenuous (low-density) component that fills
in the rest of the volume. In such a physical picture, there is no
problem for the low-density components of both massive stars in the 
binary to interact and set up the `seed' CSW region. The more massive
components of the stellar winds (the clumps) then interact with it and 
provide the strong X-ray emission observed from wide CSW binaries.

It is our understanding that the physical picture of massive stars
possessing two-component stellar winds is capable of explaining
observational characteristics of these objects along the entire
spectral range.  At least, this is the case with the CSW binary \WR
for which such a physical picture describes its optical, radio and 
X-ray properties  in a self-consistent way. We thus believe that by
adopting the same global view on other CSW binaries we can further 
check the validity of this physical picture. 
An important task for future studies is to estimate the contribution 
of the low-density component of the stellar 
wind to the observed spectrum of a massive star: an issue that needs
be addressed both from observational and theoretical point of view.

\section{Conclusions}
\label{sec:conclusions}
In this work, we presented an analysis of new optical spectra of \WR 
obtained with Gran Telescopio CANARIAS and of archive spectra from the 
Hubble Space Telescope. Our goal was to derive some basic parameters of 
both stellar components in order to check the global physical picture 
of this CSW binary.
The basic results and conclusions from our analysis of
these data are as follows.

(i) Given the spatial separation of 0\farcs64  between 
the stellar components in \WRE, individual optical spectra of its WR 
and OB components were extracted from the GTC spectra adopting a 
`deconvolution' technique, while a standard aperture extraction was
performed for the HST spectra.

(ii) The spectral modelling of the flux-calibrated WR and OB spectra 
is based on modern atmosphere models (\cmfgenE) that take into account
the optically thin clumping (volume filling factor). We adopted a 
grid-modelling approach to derive some basic physical characteristics 
of the studied massive stars: e.g., the mass-loss rate ($\dot{M}$), 
luminosity (L), stellar temperature (\TstarE). The spectral fits
provided the optical extinction E(B-V) as well.

(iii) We note that for the currently accepted distance of 630 pc to
\WRE, the values of mass-loss rate derived from modelling
its optical spectra are in acceptable correspondence with that 
from modelling its X-ray emission: \dotM $= -4.95; -4.86$~(GTC; HST)
vs. \dotM $= - 5.0$ \dotMyr (\ChandraE; \citealt{zhp_10b}).

(iv) In the light of a global view on CSW binaries, we modelled the 
radio emission from \WR as well. It was done in two ways: 
{\it directly} with \cmfgen and adopting the standard model for radio 
emission from massive stars (\citealt{pa_fe_75}; 
\citealt{wri_bar_75}). Correspondence between results from both models 
was very good, thus, showing that {\it both} can be used in analysis 
of radio emission from massive stars.
However, the mass-loss rate, derived from modelling the optical 
spectra of \WRE, gives lower radio flux than observed.  A plausible 
solution for this problem could be if the volume filling factor at 
large distance from the star (radio-formation region) is smaller 
than close to the star (optical-formation region). Adopting this, the 
model (\cmfgenE) can match well {\it both} optical and thermal radio 
emission from \WRE.

(v) The global view on the CSW binary \WR thus shows that its 
observational properties in different spectral domains 
can be explained in a self-consistent physical picture.

\section*{Acknowledgements}
Based on observations made with the Gran Telescopio Canarias (GTC),
installed in the Spanish Observatorio del Roque de los Muchachos of
the Instituto de Astrofisica de Canarias, in the island of La
Palma.This work is partly based on data obtained with the instrument
OSIRIS, built by a Consortium led by the Instituto de Astrofisica de
Canarias in collaboration with the Instituto de Astronomia of the
Universidad Autonoma de Mexico. OSIRIS was funded by GRANTECAN and the
National Plan of Astronomy and Astrophysics of the Spanish Government.
This research has made use of the NASA's Astrophysics Data System, and
the SIMBAD astronomical data base, operated by CDS at Strasbourg,
France.
The authors acknowledge financial support from
Bulgarian National Science Fund grant DH 08 12.
The authors thank an anonymous referee for 
valuable comments and suggestions.

\bibliographystyle{mnras}
\bibliography{wr147} 

\appendix
\section{Modified stellar wind clumping}
\label{app}

\begin{figure*}
\begin{center}
\includegraphics[width=2.0in, height=2.0in]{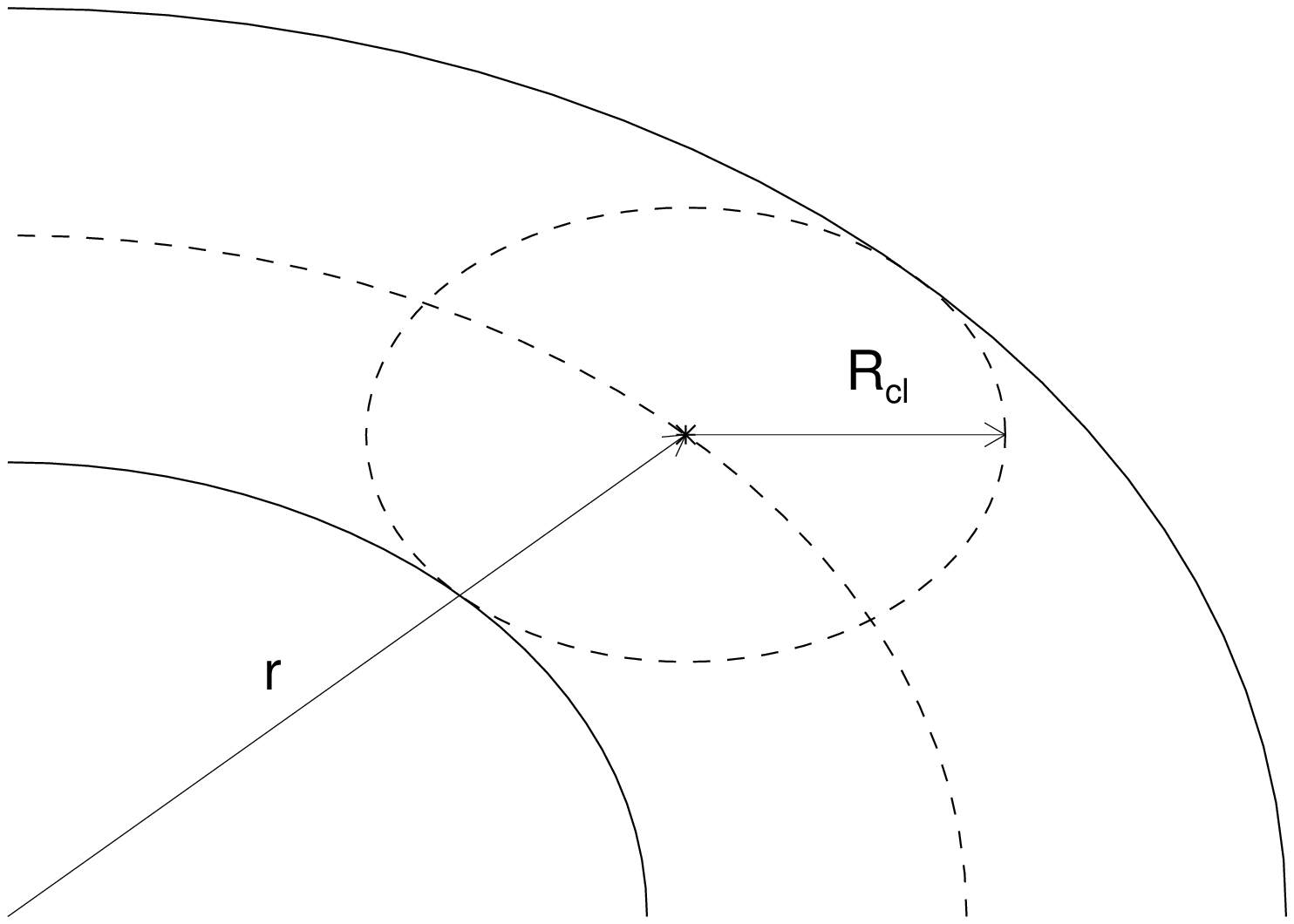}
\includegraphics[width=2.8in, height=2.0in]{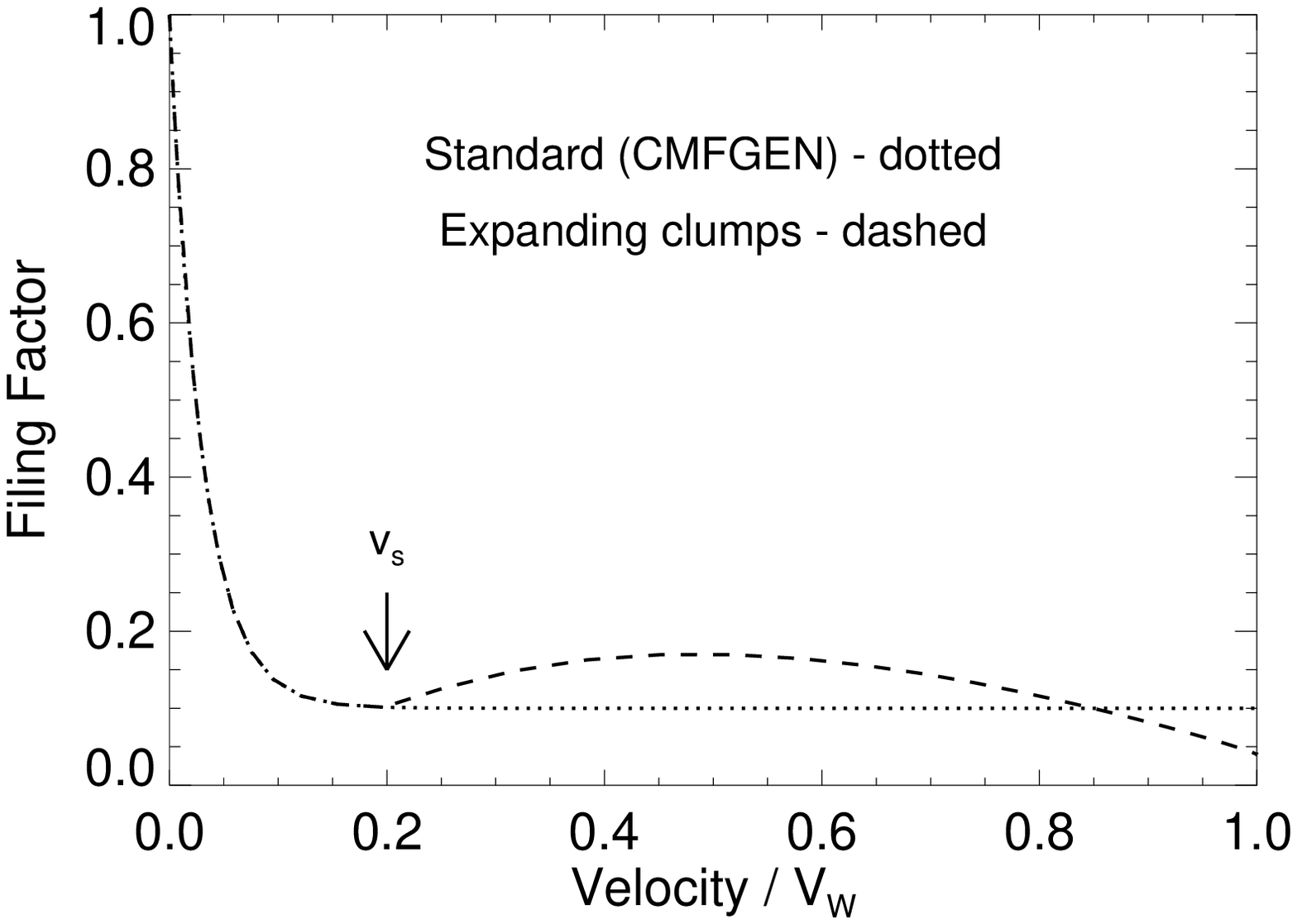}
\end{center}
\caption{The volume filling factor.
Left panel: a schematic diagram illustrating the definition of the
volume filling factor.
Right panel:  a comparison between the cases of the standard filling 
factor (CMFGEN) and that if clumps are allowed to expand.
}
\label{fig:clumping}
\end{figure*}

Let us consider a simplified picture, that is a spherically-symmetric
stellar wind is clumpy and the clumps are `identical'. So, the volume 
filling factor is defined as:
\begin{equation}
    f = \frac{\frac{4\pi}{3} R_{cl}^3 N_{cl}}{4\pi r^2 2 R_{cl}} = 
        \frac{N_{cl}}{6} \left(\frac{R_{cl}}{r}\right)^2 \nonumber
\end{equation}
where $f$ is the volume filling factor, $R_{cl}$ is the clump radius,
$r$ is the distance from the star, $N_{cl}$ is the number of clumps
(see Fig.~\ref{fig:clumping}, left panel).

Assuming that the clumps are quasi isothermal and they expand with 
the sound speed ($c_s$) determined by their plasma temperature, 
we have:
\begin{equation}
    f = \frac{N_{cl}}{6} \left[\frac{R_{cl,0} + c_s (t -
t_0)}{r}\right]^2 \nonumber
\end{equation}
where $R_{cl,0}$ is the clump radius at some initial (fiducial) time
$t_0$, which corresponds to some initial (fiducial) distance $r_0$.

For a stellar wind with a constant velocity $\varv_w$, we have:
\begin{equation}
    f = \frac{N_{cl}}{6} \left[\frac{R_{cl,0} + \frac{c_s}{\varv_{\infty}} 
        (r - r_0)}{r}\right]^2,  \hspace{1cm} 
    f_0 = \frac{N_{cl}}{6} \left(\frac{R_{cl,0}}{r_0}\right)^2,
\nonumber
\end{equation}
\begin{equation}
    f_{\infty} = f_0 \left(\frac{r_0}{R_{cl,0}}\right)^2
\left(\frac{c_s}{\varv_w}\right)^2 \nonumber
\end{equation}
where $f_0$ is the volume filling factor at radius $r_0$ and
$f_{\infty}$ is the filling factor at large radius (infinity).

For a stellar wind with a $\beta$-law velocity
($\beta = 1$), we have:
\begin{equation}
    \varv = \varv_0 + (\varv_w - \varv_0) \left(1 - \frac{r_0}{r}\right),
\hspace{1cm}
    \varv = \varv_w \left[ \alpha + (1 - \alpha) \left(1 -
\frac{r_0}{r}\right)\right] \nonumber
\end{equation}
where $\varv_0$ is the wind velocity at radius $r_0$, $\varv_{\infty}$ is
the terminal wind velocity and $\alpha = \varv_0 / \varv_w$  

So, we can write:
\begin{equation}
    t - t_0 = \int dt = \int \frac{dr}{\varv} = 
            \frac{r_0}{\varv_{\infty}} \int \frac{d\rr}{ \alpha + 
            (1 - \alpha) \left(1 - \frac{1}{\rr}\right)}  \nonumber
\end{equation}
as the distance $r$ is normalized to $r_0$, that is at time $t_0$ the
distance is $\rr = 1$ ($\rr = r / r_0$). 
After some manipulations we end up with:
\begin{equation}
    t - t_0 = \frac{r_0}{\varv_{\infty}} 
              \left[ (\rr - 1) + (1 - \alpha) \ln\frac{\rr - (1 -
              \alpha)}{\alpha}\right] \nonumber
\end{equation}

And for the filling factor, we have:
\begin{equation}
    f = f_0 \left\{
        \frac{1 + \frac{r_0}{R_{cl,0}} \frac{c_s}{\varv_{\infty}}
              \left[ (\rr - 1) + (1 - \alpha) \ln\frac{\rr - (1 -
              \alpha)}{\alpha}\right]}{\rr}
            \right\}^2 \nonumber
\end{equation}
or
\begin{equation}
    f = f_0 \left\{
        \frac{1 + \sqrt{\frac{f_{\infty}}{f_0}} 
              \left[ (\rr - 1) + (1 - \alpha) \ln\frac{\rr - (1 -
              \alpha)}{\alpha}\right]}{\rr}
            \right\}^2 \nonumber
\end{equation}

Note, the volume filling factor at large distances could be 
smaller or larger than its initial value, for example:
\begin{equation}
if\,\,\,    f_{\infty} < f_0 
             \hspace{0.2cm} \Longrightarrow \hspace{0.2cm} 
\left(\frac{r_0}{R_{cl,0}}\right)^2 \left(\frac{c_s}{\varv_w}\right)^2 < 1
             \hspace{0.2cm} \Longrightarrow \hspace{0.2cm}  
   \frac{R_{cl,0}}{r_0} > \frac{c_s}{\varv_w} \nonumber
\end{equation}

We could assume that clumps form at distances smaller than
some `setup' distance/radius ($r_s > r_0$), so, the filling factor
has its standard \cmfgen form (see eq.~\ref{eqn:clumping}) in this
inner region ($r < r_s$).  And, beyond that clump-formation region 
($r > r_s$) clumps may only expand but no new clumps appear. In such a
case, the clump expansion will start at some initial time $t_s$ and
similarly as above, we can derive:
\begin{eqnarray}
    f &=& f_s \left\{
        \frac{\rr_s + \sqrt{\frac{f_{\infty}}{f_0}} 
              \left[ (\rr - \rr_s) + (1 - \alpha) \ln\frac{\rr - (1 -
              \alpha)}{\rr_s - (1 - \alpha})\right]}{\rr}
            \right\}^2   \nonumber \\
  & &  \\
    f_s &=& \frac{N_{cl}}{6} \left(\frac{R_{cl,0}}{r_s}\right)^2,
\hspace{1cm}
    f_{\infty} = f_s \left(\frac{r_s}{R_{cl,0}}\right)^2
\left(\frac{c_s}{\varv_w}\right)^2  \nonumber
\end{eqnarray}

We can also express the volume filling factor in terms of flow
velocity with $\varv_s$ denoting its value at the clump-formation
distance/radius $r_s$ ($\vv = \varv / \varv_{\infty}$):
\begin{eqnarray}
   & &f = \frac{f_s}{(1-\vv_s)^2}  \times  \nonumber  \\
   & &   \\
   & &\left\{
        1 - \vv + \sqrt{\frac{f_{\infty}}{f_s}}
        \left[ \vv - \vv_s + (1 - \vv)(1 - \vv_s)\ln\left(\frac{\vv}{1 -
        \vv}\frac{1 - \vv_s}{\vv_s}\right)\right]
            \right\}^2  \nonumber
\end{eqnarray}

An example of how the filling factor changes if the clumps are allowed
to expand is shown in Fig.~\ref{fig:clumping} (right panel). The
standard filling factor (CMFGEN; see eq.~\ref{eqn:clumping}) is 0.1 
while the case of expanding
clumps has a value of $f_{\infty} = 0.04$ at large radii from the star 
(where the wind reaches the terminal wind velocity). In this
particular case, the clump-formation radius corresponds to 
$\vv_s \approx 0.2$.

\section{Model atoms used in \cmfgen}
\label{app_atom}

Adopted atomic data of all elements included in our model atmosphere 
calculations are summarized in Table~\ref{tab:atom}. While the energy 
levels for HeI are from \citet{martin_87},
most of the atomic data are from the Opacity \citep{seaton_87} and the 
Iron \citep{hummer_93} projects. However, for some CNO elements, 
atomic data were used also from \citet{nussbaumer_83,nussbaumer_84},
whilst for Iron  data were used  from \citet{nahar_95},
\citet{zhang_96}, \citet{becker_95a,becker_95b}.


      \begin{table}
	\caption{Model atoms used in our models. For each ion the number 
         of full levels, super levels, and bound-bound transitions 
         is provided.\label{tab:atom}}%
	\begin{center}
	  \tabcolsep1.8mm
	  \begin{tabular}{l r r r } 
	  \hline  
	  \hline
	  Ion & Super levels   & Full levels & b-b transitions \\
	  \hline
	  H  I      &   20 &  30  &   435   \\ 
	  He I      &   45 &  69  &   907   \\ 
	  He II     &   22 &  30  &   435   \\ 
	  C  II     &   10 &  18  &    47   \\
	  C  III    &   99 & 243  &  5528   \\ 
	  C  IV     &   59 &  64  &  1446   \\ 
	  C   V     &   46 &  73  &   490   \\ 	  
	  N  II     &    9 &  17  &    29   \\ 
	  N  III    &   41 &  82  &   582   \\ 
	  N  IV     &  200 & 278  &  6943   \\ 
	  O  II	    &   81 & 182  &  2800   \\ 
	  O  III    &  165 & 343  &  6516   \\ 
	  O  IV	    &   71 & 138  &  1597   \\ 

	  Ne II     &   25 & 116  &  1732   \\ 
	  Ne III    &   57 & 188  &  2343   \\ 
	  Mg III    &   41 & 201  &  3052   \\ 

	  Al III    &    7 &  12  &    26   \\ 
	  Al  IV    &   62 & 199  &  2992   \\

	  Si III	&   26 &   51 &   245   \\ 
	  Si  IV    &   55 &   66 &  1090   \\ 
	  Si   V    &   52 &  203 &  3086   \\ 	  
	  P  IV	    &   30 &   90 &   656   \\  
	  P  V      &    9 &   15 &    42   \\ 
	  S  III    &   41 &   83 &   601   \\ 
	  S  IV	    &   69 &  194 &  3598   \\ 
	  S  V 	    &   41 &  167 &  2229   \\

	  Ar III    &   18 &   82 &   604   \\
	  Ar IV 	&   41 &  204 &  3459   \\

	  Ca III    &   41 &  208 &  2784   \\
	  Ca IV 	&   39 &  341 &  7250   \\	  

	  Fe IV	    &  100 & 1000 & 37899   \\ 
	  Fe V	    &   45 &  869 & 31688   \\ 
	  Fe VI	    &   55 &  674 & 20240   \\ 
	  Fe VII	&   13 &   50 &   237   \\ \hline
	  \end{tabular}
	  \label{table:soph_models} 
      \end{center}
      \end{table}

\bsp    
\label{lastpage}
\end{document}